\documentclass[%
reprint,
 amsmath,amssymb,
 aps,
 pra,
]{revtex4-2}

\usepackage{graphicx}
\usepackage{dcolumn}
\usepackage{bm}
\usepackage{color}
\usepackage{array}
\usepackage{ulem}

\newcommand{\kone}{\kappa_{1\text{ph}}}
\newcommand{\ktwo}{\kappa_{2\text{ph}}}
\newcommand{\epstwo}{\epsilon_{2\text{ph}}}
\newcommand{\ha}{\hat{a}}

\newcommand{\hb}{\hat{b}}

\newcommand{\hatwo}{\hat{a}^2}
\newcommand{\hadtwo}{\hat{a}^{\dagger 2}}

\newcommand{\ket}[1]{|#1\rangle}

\newcommand{\mm}[1]{\textcolor{black}{#1}}
\newcommand{\jg}[1]{\textcolor{black}{#1}}

\begin{document}

\title{Error Rates and Resource Overheads of Repetition Cat Qubits}

\author{Jérémie Guillaud}
\email{jeremie.guillaud@gmail.com}
\author{Mazyar Mirrahimi}%
\affiliation{%
QUANTIC team, Inria Paris, 2 rue Simone Iff, 75012 Paris, France.
}%

\date{\today}

\begin{abstract}
We estimate and analyze the error rates and the resource overheads of the repetition cat qubit approach to universal and fault-tolerant quantum computation. The cat qubits stabilized by two-photon dissipation exhibit an extremely biased noise where the bit-flip error rate is exponentially suppressed with the mean number of photons. In a recent work, we suggested that the remaining phase-flip error channel could be suppressed using a 1D repetition code. Indeed, using only bias-preserving gates on the cat qubits, it is possible to build a universal set of fault-tolerant logical gates at the level of the repetition cat qubit. In this paper, we perform Monte-Carlo simulations of all the circuits implementing the protected logical gates, using a circuit-level error model. Furthermore, we analyze two different approaches to implement a fault-tolerant Toffoli gate on repetition cat qubits. These numerical simulations indicate that very low logical error rates could be achieved with a reasonable resource overhead, and with  parameters that are within the reach of near-term circuit QED experiments.
\end{abstract}

\maketitle


\section{\label{sec:introduction}Introduction}
The theory of quantum fault-tolerance~\cite{Shor1995,Shor1996,Steane1996} ensures that arbitrarily long quantum computations can be reliably performed  on a noisy quantum hardware. However, this comes at the price of a tremendous hardware overhead, such that a big focus of the last few decades of research in quantum information theory has been to develop hardware-efficient protocols for universal and fault-tolerant quantum computation~\cite{Campbell2017}. In a recent attempt to reduce this physical resource overhead, we proposed to encode quantum information in a 1D repetition code based on cat qubits~\cite{Guillaud2019}. In the large and actively studied family of bosonic quantum codes~\cite{Joshi2020}, the cat qubits' specific encoding stands out by the remarkable property that any error process acting \mm{locally in the phase-space of the bosonic mode,} mostly produces phase-flip errors in the two-dimensional manifold defining the cat qubit \jg{\cite{Cochrane1999, Ralph2003, Mirrahimi2014}}. More rigorously, given an error process of the bosonic mode (say, photon loss) occurring at a rate $\kappa$, and a cat qubit parametrized by a complex amplitude $\alpha \in \mathbb{C}$ and a two-photon dissipation rate $\kappa_{2\text{ph}}$ larger than  $\kappa$, the effective rate of resulting bit-flips acting on the cat qubit is suppressed exponentially in the average number of photons $\bar n = |\alpha|^2$, while the rate of phase-flips increases linearly~\cite{Lescanne2020}. 
Previous works have investigated how to leverage the noise bias of such qubits by adapting appropriately the error correcting codes in order to improve the overall performance of the scheme~\cite{Aliferis2008,Tuckett2018,Tuckett2020, bonillaataides2020}. Here, we take one step further and anticipate that by increasing the average number of photons, the noise bias in the dissipative cat qubits can be made so important that the error suppression provided by the stabilizer code needs only to suppress the remaining dominant error. Indeed, the scaling of the resulting error rates for the encoded cat qubit is somewhat similar to the protection achieved by a 1D repetition code protecting against bit-flips, where the role of the distance $d$ of the code is now played by the average number of photons $\bar n$. Interestingly, this inner code 'distance' can be augmented without increasing the number of physical quantum systems. In the spirit of Bacon-Shor codes
~\cite{Shor1995,Bacon2006}, the full protection of the quantum bit of information is achieved by embedding the cat qubit into a dual repetition code protecting against phase-flips, the so-called repetition cat qubit. Because the family of Bacon-Shor codes does not possess an accuracy threshold, these codes cannot be used alone to achieve arbitrarily low error rates. Rather, in many constructions, a large block of these codes is used at the bottom level to obtain very low logical error rates. Then, to reach arbitrarily low logical error rates, the code needs to be concatenated with any other code possessing a threshold
~\cite{Cross2009,Aliferis2007} . The interest of using a Bacon-Shor code at the bottom level, in these constructions, is usually that it reduces the overall resource overhead, or to gain access to new logical operations.

However, even without a threshold, it has been shown that a single block of Bacon-Shor code used without concatenation could achieve extremely low error rates at the cost of a reasonable resource overhead~\cite{Brooks2013,Cross2009,Napp2013}. For this reason, Bacon-Shor codes and related constructions are promising candidates for near-term experimental demonstration of quantum error correction and fault-tolerant processing of encoded quantum information. As we will see in this paper, the repetition cat qubit approach is similar: even though strict fault-tolerance requires concatenating the repetition cat qubit with a code that possesses an accuracy threshold, a large enough repetition cat qubit alone already yields extremely low logical error rates with a reasonable resource overhead. This level of protection should allow us to perform useful quantum computations without resorting to code concatenation.

The rigorous characterization of the performance of a given architecture is a tricky task. The obtained performance is highly dependent on the underlying assumptions made about the noise structure, the quantum computer's architecture, and the types of gates that are considered. In this paper, we rely on a circuit-based error model for the  simulations of the operations in our universal set, both Clifford and non-Clifford. These simulations provide realistic estimates of the expected logical error rates and resource overheads, using physical parameters that are within the reach of the circuit QED experiments in the next few years.

The paper is structured as follows. We recall the repetition cat qubit approach to universal and fault-tolerant quantum computation in Section~\ref{sec:prx}. The methodology and the assumptions used to derive the error rates and overhead are discussed in  Section~\ref{sec:Assumptions and Methodology}. In Section~\ref{sec:CliffordOperations}, we estimate the error rate and resource overhead of a repetition cat qubit used as a quantum memory, and of the logical gadgets that admit a transversal construction. We then discuss in Section
~\ref{sec:Toffoli} the implementation of the non-Clifford resource needed for universality: the Toffoli gate, and give estimation of its performance. In Section~\ref{sec:architecture}, we start a discussion around the question of architecture of a quantum processor based on repetition cat qubits. The intention of this discussion is to suggest a few possible research directions rather than providing a final and optimal solution. \jg{We note that during the reviewing process of this manuscript, a comprehensive analysis of the performance of a concatenated cat qubit architecture became available \cite{Chamberland2020_AWS}. This work completes our own, and provides new numerical and analytical tools to analyse the specific dynamics of cat qubits.}

\section{\label{sec:prx} Repetition cat qubits for fault-tolerant quantum computation}
This section summarizes the key results of~\cite{Guillaud2019}. The starting point of the approach is the two-photon driven dissipative process that stabilizes cat qubits, and provides an autonomous protection against bit-flips. The suppression of the remaining phase-flip errors is achieved with a 1D repetition code. Then, we recall the set of gates or operations acting on the cat qubit that are compatible with the exponential suppression of bit-flip errors. The operations in this set are combined to design gadgets acting as logical gates on the repetition cat qubit.

A complete theoretical presentation of the two-photon pumped cat qubits can be found in~\cite{Mirrahimi2014}, and experimental demonstrations of this proposal in~\cite{Leghtas2015, Touzard2018, Lescanne2020}. The two-photon driven dissipative process consists in applying a two-photon drive of amplitude $\epsilon_{2\text{ph}}$ to a mode that can only exchange photons in pairs with its environment at a rate $\ktwo$. The Lindblad master equation describing this dynamics is:
\begin{equation}
    \dfrac{d\rho}{dt} = [\epstwo \hadtwo - \epstwo^*\hatwo, \rho] + \ktwo \mathcal{D}[\hatwo]\rho = \ktwo \mathcal{D}[\hatwo - \alpha^2] \rho
\end{equation}
where $\mathcal{D}[\hat L] \rho = \hat L \rho \hat L^\dagger - \tfrac12\hat L ^\dagger \hat L \rho - \tfrac12 \rho \hat L ^\dagger \hat L$ and $\alpha = \sqrt{2 \epstwo / \ktwo}$. This dynamics stabilizes the two-dimensional manifold spanned by the coherent states $\ket{\alpha}$ and $\ket{-\alpha}$, or equivalently, by the coherent superpositions of these states, known as Schrödinger cat states:
\[
    \ket{\mathcal{C}_\alpha^\pm} := \mathcal{N}_\pm (\ket{\alpha} \pm \ket{-\alpha}) 
\]
where $\mathcal{N}_\pm = (2(1\pm\exp(-2|\alpha|^2))^{-1/2}$ is a normalization factor. Following the convention of~\cite{Guillaud2019, Lescanne2020}, the cat states $\ket{\mathcal{C}_\alpha^\pm}$ are chosen to be the $\pm 1$ eigenvectors of the Pauli X operator for the cat qubit, and the computational basis is defined as:
\begin{align*}
    \ket{0}_c &= (\ket{\mathcal{C}_\alpha^+} + \ket{\mathcal{C}_\alpha^-})/\sqrt2 = \ket{\alpha} + \mathcal{O}(e^{-2|\alpha|^2}), \\
    \ket{1}_c &= (\ket{\mathcal{C}_\alpha^+} - \ket{\mathcal{C}_\alpha^-})/\sqrt2 = \ket{-\alpha} + \mathcal{O}(e^{-2|\alpha|^2}).
\end{align*}


The non-locality of information in the phase space ensures an exponential suppression of bit-flips with the cat size $|\alpha|^2$~\cite{Lescanne2020}. This raises the hope that a simple quantum error correcting code correcting phase-flips only could be sufficient to achieve fault-tolerance. The simplest choice is a repetition code in the dual basis. The fully protected logical qubit obtained is called the \textit{repetition cat qubit}. For a distance $d$ code encoding one repetition cat qubit into $d$ cat qubits, the $d-1$ stabilizers are $X_i X_{i+1}$, $i \in [\![ 0, d-1 ]\!]$ and the logical Pauli operators are $X_L = X_k$ ($k \in [\![ 0, d-1 ]\!]$), $Z_L = \bigotimes Z_i$ and $Y_L = -i Z_L X_L$.

We now describe how operations can be performed on a cat qubit. The manipulation of the quantum information encoded in the cat qubit requires special care to preserve the noise structure. More precisely, to carry through the exponential suppression of the bit-flip error rate, any operation performed on the cat qubit has to be \textit{bias-preserving}, which means two things: it must not convert a phase-flip error into a bit-flip error, and it must preserve the exponential suppression of bit-flips while the operation is being performed. Note that the first requirement automatically rules out certain gates, such as the Hadamard gate $H = (X + Z)/\sqrt 2$ (not allowed because $HZ = XH$). The second requirement is  satisfied if the non-locality of the information in the phase space is maintained during the execution of the gate. More precisely, given a unitary $U$ on the cat qubit implemented by the dynamics $\dot{\rho} = \mathcal{L}(\rho)$ in time T, $(U = e
^{T\mathcal{L}})$, the noise bias is preserved during the execution of the gate $U$, if the two states $e^{t\mathcal{L}}(\ket{\pm\alpha})$ remain distant in the phase space for all times  $t \in [0, T]$.

Fortunately, an important class of operations can be performed in this way \jg{\cite{Puri2020, Guillaud2019}}. This class includes the preparation of the eigenstates $\ket{\pm}_c$ of Pauli operator X (noted as $\mathcal{P}_{\ket{\pm}_c}$), the measurement of the observable X  (noted as $\mathcal{M}_X$),  single qubit X logical gate, rotation of an arbitrary angle $\theta$ around the Z axis $Z(\theta) = \exp(-i \frac{\theta}{2} Z)$,  two-qubit controlled-$X$ gate (also called CNOT and denoted CX) and the controlled-$Z(\theta)$ gate, and finally the three-qubit controlled-controlled-$X$ gate (also called Toffoli and denoted CCX). Performing a CNOT or a Toffoli gate with a bias-preserving process is non trivial \jg{\cite{Puri2020}}. The extra degree of freedom associated with the choice of the complex parameter $\alpha$ defining the cat qubit plays a crucial role in the design of bias-preserving processes: more precisely, it has been shown that for a conventional two-level system, the CNOT or Toffoli gates cannot be implemented in a bias-preserving manner
~\cite{Guillaud2019}. For this reason, in previous work on quantum computation with biased-noise qubits~\cite{Aliferis2008}, these gates were usually discarded. To clarify the important role of the underlying Hilbert space of the harmonic oscillator in the bias-preserving implementations of gates, we recall how the X gate and the CNOT gate are realized on cat qubits. The  physical realizations of the other above operations are described in~\cite{Guillaud2019}.

The X gate, swapping the states $\ket{0}_c\approx\ket\alpha$ and $\ket{1}_c\approx\ket{-\alpha}$, is realized in time $T$ by adiabatically modulating the phase of the two photon drive $\epstwo \rightarrow \epstwo e^{2i\frac{\pi}{T} t}$, effectively transforming the two-photon dissipator $\mathcal{D}[\ha^{2} - \alpha^2]$ to $\mathcal{D}[\ha^{2} - (\alpha e^{i\frac{\pi}{T}t})^2]$. The instantaneous eigenstates of this time-dependent superoperator are $\ket{\pm \alpha e^{i\frac{\pi}{T}t}}$, taking an initial state $\ket{\psi(t=0)} = c_0 \ket\alpha + c_1 \ket{-\alpha}$ to the state $\ket{\psi(t)} = c_0 \ket{\alpha e^{i\frac{\pi}{T}t}}+ c_1 \ket{-\alpha e^{i\frac{\pi}{T}t}}$ after a time t, resulting in a final state $\ket{\psi(t=T)} = c_0 \ket{-\alpha} + c_1 \ket{\alpha} = X \ket{\psi(t=0)}$. 

In order to extend the above single-qubit X gate to a two-qubit controlled-X gate, we must perform the  continuous evolution of the dissipator $\mathcal{D}[\ha^{2} - (\alpha e^{i\frac{\pi}{T}t})^2]$ on the target cat qubit (stabilized in the mode $\ha$) only when the control cat qubit (in the mode $\hb$) is in the state $\ket{-\alpha}$. Otherwise, the dissipator $\mathcal{D}[\ha^{2} - \alpha^2]$ needs to be retained. This can be done through the time-dependent dissipator $\mathcal{D}[\ha^2 - \tfrac12\alpha(\hb+\alpha) + \tfrac12\alpha {e^{2 i \frac \pi T t}}(\hb-\alpha)]$ and the constant one $\mathcal{D}[\hb^2 - \alpha^2]$. During  the dynamics the instantaneous steady states of the dissipators remain distant in the 4D phase space of the two harmonic oscillators and therefore the exponential bit-flip suppression is maintained. Some amount of phase-flip errors is induced by the finite adiabaticity of the gate. As shown in~\cite{Guillaud2019}, these non-adiabatic effects can be reduced by using a feed-forward Hamiltonian 
$$
H=\frac{\pi}{2T}\frac{(a-\alpha)}{2\alpha}\otimes(b^\dag b-|\alpha|^2)+\text{h.c.}
$$
In presence of the above dissipators and feed-forward Hamiltonian, increasing the gate time $T$ further reduces the phase-flip error probability due to non-adiabatic effects but increases the probability that they are induced by other noise channels such as single photon loss. The gate time $T^*$ that maximizes the CNOT fidelity in presence of single photon loss (occurring at a rate $\kone$) is given by $T^* = (2\bar n \sqrt\pi)^{-1}\sqrt{1/\kone\ktwo}$~\cite{Guillaud2019}.

\begin{figure}[h]
\includegraphics[width=.25\textwidth]{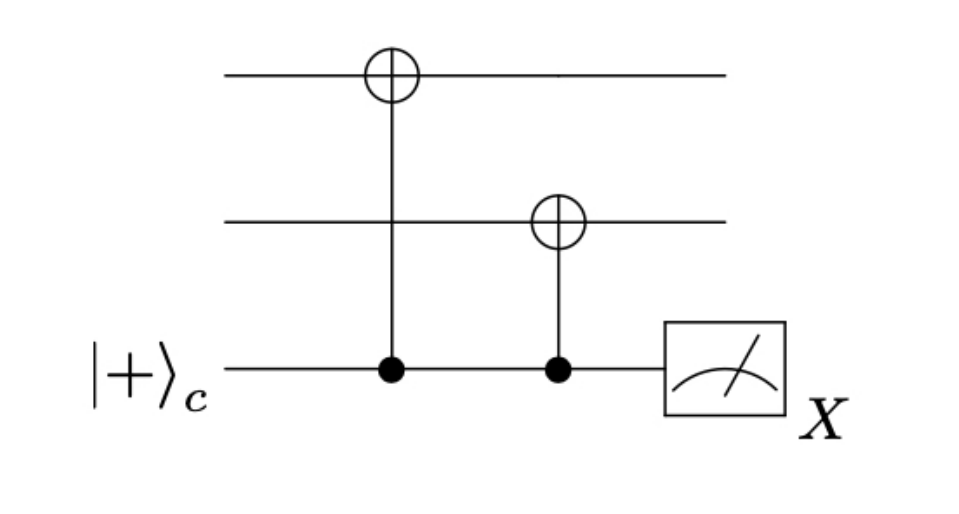}
\caption{Quantum circuit to measure a $X_1 \otimes X_2$ stabilizer of the repetition code (top two qubits). Every operation in this circuit can be done in a bias-preserving manner. \label{fig:stabilizer_measurement_circuit}}
\end{figure}

Before attempting to build logical operations on the repetition cat qubit, an important issue is whether the stabilizers can be measured in a bias-preserving manner. A standard circuit used to perform a quantum non demolition measurement of the stabilizers of a repetition code is depicted in Fig.~\ref{fig:stabilizer_measurement_circuit}. Since every operation in this circuit can be done in a bias-preserving manner, the circuit itself is bias-preserving. There are other protocols that could equivalently be used for the same purpose, see e.g
~\cite{Puri2019}.

\begin{figure*}[t!]
\includegraphics[width=.9\textwidth]{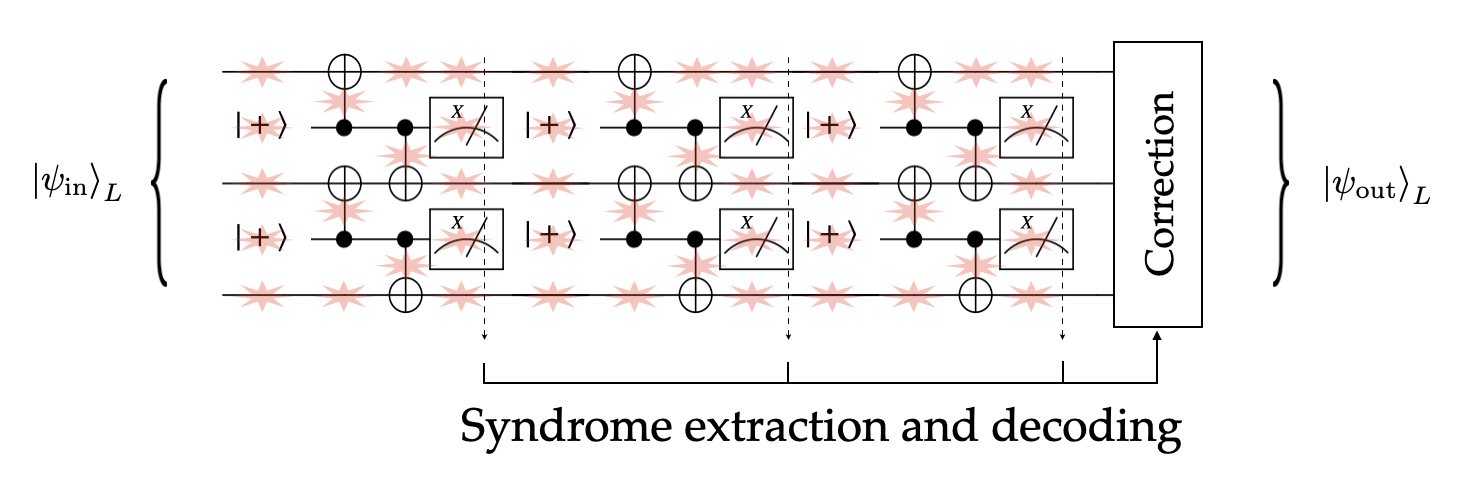}
\caption{Quantum error correction for a repetition cat qubit used as a quantum memory (here the code distance $d = 3$). The $d-1$ stabilizers are measured $d$ times, and the measurement results are used in a MWPM decoder. The red stars mark the possible locations of errors.
\label{fig:QEC_circuit}}
\end{figure*}

\section{\label{sec:Assumptions and Methodology} Assumptions and Methodology}

\subsection{\label{subsec:pseudo-threshold}Assumptions}
The repetition cat qubit approach to universal and fault-tolerant quantum computation relies on two different kinds of protection. The two-photon dissipation exponentially suppresses bit-flip errors with the mean number of photons in the cat state, while the rate of phase-flip errors increases only linearly. Next, the repetition code suppresses exponentially the phase-flip errors, provided that the phase-flip error rate of the cat qubit is below the fault-tolerance threshold of the repetition code.

This protection is similar to the one achieved by Bacon-Shor codes~\cite{Shor1995,Bacon2006}, with the nice feature that the 
``distance'' of the inner protection provided by the two-photon pumping can be increased without any further hardware overhead. However, similarly to Bacon-Shor codes, because the phase-flip error rate of the cat qubit increases linearly with the mean number of photon, there cannot be a threshold since the effective phase-flip error rate of the cat qubit will eventually exceed the threshold of the repetition code. Nonetheless, this is not an obstacle to obtaining extremely low logical error rates with this approach, by limiting the mean number of photons to a finite value for which the bit-flip error probability is extremely low, and for which the phase-flip error probability is still below the threshold of the repetition code. In our analysis, we limit the size of the cat qubits to \jg{$\bar n = 15$ photons} and assume that the exponential suppression of the bit-flip error rate holds at least up to this cat size. 

All the circuits presented in this work are built to implement logical gates on repetition cat qubits, while being fault-tolerant to phase-flip errors only. Indeed, any single bit-flip error occurring on any data qubit during the execution of a circuit can cause a logical bit-flip errors. The resulting logical bit-flip error rate can therefore be bounded by simply counting the number of single locations in the circuits where a bit-flip can occur. The numerical simulations of the logical circuits is devoted to estimating the logical phase-flip error rate only, without taking into account the bit-flip errors. For fault-tolerant circuits with respect to phase-flip errors, we define the ``phase-flip threshold'' to be the highest value of the physical phase-flip error probability $p_{th}$ for which the logical phase-flip error probability decreases upon an increase of the code distance $d$.

\subsection{\label{subsec:fault-tolerant simulation}Fault-tolerant simulation of a quantum circuit}

The logical error probability of the different logical gates in our universal gate set are numerically estimated by performing Monte Carlo simulations of the circuits implementing these gates. We consider a circuit-based error model which takes into account errors of all operations in the circuit, including identity for the idle qubits. This is a more realistic model than the so-called phenomenological error model, \jg{where the operations of the syndrome extraction circuit are assumed to be perfect and the syndrome measurements are assigned an error probability modelled phenomenologically.} \mm{In contrast, the circuit-based error model takes into account the errors of all operations in the circuit, and is best suited for estimating the logical error rate in the context of quantum computation.}

The error models for various physical operations of Fig.~\ref{fig:QEC_circuit} are provided in~\cite{Guillaud2019} and summarized in Table~\ref{tab:error_models}. While the bit-flip error probability remains exponentially suppressed in $\bar n$, applying a noisy CNOT gate, for instance, consists in applying a perfect CNOT gate, followed by a probabilistic application of either identity with probability $1 - 4p$, or a Z error on the control qubit with probability $3p$, or a Z error on the target qubit with probability $p/2$, or a correlated Z error on both qubits with probability $p/2$. Here the parameter $p$ is given by $\bar n\kone T$  which for the optimal gate time $T^*$ is given by
\begin{equation}\label{eq:probaOpt}
p=(2\sqrt\pi)^{-1}\sqrt{\kone/\ktwo}.
\end{equation}

Each circuit is divided into time-steps, where every qubit (both data and ancilla) in the circuit is acted upon at every time step. For simplicity, we fix the duration of these time-steps to be the same as $T$, the duration of CNOT and Toffoli gates. When a qubit is acted upon by a gate at a given time-step, the applied probabilistic error is drawn from the corresponding error model, otherwise, the error is drawn from the identity error model (which corresponds to a phase-flip probability of $p=\bar n \kone T$).

In order to simulate a given circuit, we fix the code distance $d$ and the value of the physical noise strength $p$ and run the noisy circuit $N$ times. For each trajectory, the output of the syndrome measurements is decoded using a minimum weight perfect matching (MWPM) decoder and a final perfect recovery operation~\cite{Fowler2012b}. After the recovery operation, mapping the state back to the codespace, we check whether a logical error has occurred. We then define the logical error probability of the circuit $p_{L}(d,p)$ as
$$
p_{L} = \dfrac{N_{\text{fail}}}{N}
$$
where $N_{\text{fail}}$ is the number of times a logical error occurred during the N runs. All the circuits simulated in this work are run continuously until at least $N_{\text{fail}} = 500$ logical failures are observed, which ensures that the relative error on $p_{L}$ is less than $9\%$ with probability $95\%$. The decoding step is computed efficiently using Dijkstra's shortest path algorithm~\cite{Dijkstra1959} to generate the graph of detection events and the minimum weight perfect matching algorithm to match these events~\cite{edmonds_1965,Kolmogorov2009}. The simulations of the non-Clifford circuits of Section~\ref{sec:Toffoli} are slightly more complicated. In general, classical simulations of non-Clifford circuits become rapidly intractable with large distances $d$. However, the particular structure of the circuit and the bias-preserving property of the physical gates makes it possible to efficiently simulate the propagation of errors throughout the circuit. More precisely, the Pauli phase-flip errors either remain Pauli or propagate to a Clifford CZ type error which commutes with the rest of the circuit, up until a point where it meets a measurement that eventually projects it. Importantly, the errors can never propagate to non-Clifford errors. This feature, which is particular to our circuits and error models, is not true in all generality for random non-Clifford circuits, but allows us to numerically simulate the logical error probability for our non-Clifford circuits using only the CHP algorithm introduced in~\cite{Aaronson2004}.  We refer the reader interested in these implementation considerations to Appendix \ref{appendixA}. The numerical computations were performed in parallel using the cluster of Inria Paris, composed of 68 nodes for a total of 1244 cores. The nodes are divided in a few hardware generations: 28 bi-processors Intel Xeon X5650 of 6 cores, 12 bi-processors E5-2650v4 2.20 of 12 cores, 16 bi-processors XeonE5-2670 of 10 cores, 8 bi-processors E5-2695 v4 of 18 cores, 4 bi-processors E5-2695 v3 of 14 cores. Some data points for the logical Toffoli circuits corresponding to the largest distances and lowest logical error probabilities, for which $\sim 10^8 - 10^9$ trajectories were simulated per point, required up to a week (real time) of computation.

\begin{table*}[t!]
\begin{ruledtabular}
\begin{tabular}{|cc|cc|cc|cc|cc|cc|}
 \multicolumn{2}{|c|}{I} & \multicolumn{2}{c|}{$\mathcal{P}_{\ket+}$} & \multicolumn{2}{c|}{Z} & \multicolumn{2}{c|}{CZ} & \multicolumn{2}{c|}{CNOT} & \multicolumn{2}{c|}{Toffoli} \\ \hline
Error & Probability & Error & Probability & Error & Probability & Error & Probability & Error & Probability & Error & Probability \\ \hline
$I$ & $1-p$ & $I$ & $1-p$ & $I$ & $1-p$ & $I$ & $1-2p$ & $I$ &  $1-4p$ & $I$ &  $1-6p$ \\
$Z$ &  $p$ & $Z$ & $p$ & $Z$ & $p$ & $Z_1$ & $p$ & $Z_1$ & $3p$ & $Z_1$ & $p$ \\
 & &    &     &     &     & $Z_2$ & $p$ & $Z_2$ & $p/2$  & $Z_2$ & $p$ \\
 & &   &     &     &     &       &     & $Z_1 Z_2$ & $p/2$  & $Z_3$ & $p/2$ \\
& & &    &     &     &       &     &  &   & $\text{CZ}_{12}$ & $3p$ \\
& & &     &     &     &       &     &  &   & $\text{CZ}_{12}Z_3$ & $p/2$ \\
\end{tabular}
\end{ruledtabular}
\caption{\label{tab:error_models}Error models of each gate used in the simulations. Every noisy gate is modelled as a perfect gate, followed by a stochastic error. For each gate, we summarize the Z-type errors and the corresponding probability that we have used in the Monte Carlo simulations. We also assume the ancilla measurement to be faulty with probability $p$. These error models account for non-adiabatic errors and for the effect of single photon loss at rate $\kone$. The parameter $p$ that characterizes the ``strength'' of the physical noise is the error probability of a physical phase-flip during the typical gate time $T$, given by $p = \bar n \kone T$. For the gate time $T
^*$ that maximizes the CNOT and Toffoli gate fidelities, this probability is given by $p = \frac{1}{2\sqrt{\pi}}\sqrt{\kone/\ktwo}$. For simplicity sakes, we assume that all the gate times are equal.
}
\end{table*}

\section{\label{sec:CliffordOperations} Memory and transversal Clifford operations: performance}

\subsection{\label{subsec:memory} Repetition cat qubit as a quantum memory}

In this subsection, we investigate the performance of a repetition cat qubit used as a quantum memory. The QEC is applied to extend the lifetime of a quantum bit of information. In this case, the logical circuit (implementing the logical encoded version of identity operation) simply consists of the error correction step. The $d-1$ stabilizer operators are measured using $d-1$ ancilla qubits. To make the procedure fault-tolerant, the measurements are repeated $d$ times before they are decoded with a MWPM decoder. Since all the operations are bias-preserving and do not convert $X$ and $Z$ errors, we can separately estimate the error probabilities $p_{Z_L}$ and $p_{X_L}$ of  logical $Z_L$ and $X_L$ errors occurring per cycle of error correction. Then, the logical error probability is bounded by $p_L = p_{Z_L} + p_{X_L}$.

\paragraph{Logical X error probability $p_{X_L}$}
\jg{
The repetition code does not provide any protection against bit-flip errors. Hence, a single bit-flip occurring on any qubit during the execution of the circuit will cause a logical $X_L$ error. While the bit-flip rate is exponentially suppressed with the mean number of photons in the cat state (experimentally observed in \cite{Lescanne2020} for the identity operation), the exact value of this probability depends on the operation that is applied to the cat qubit.} \\

\jg{The value of the bit-flip error probability induced by each gate is determined numerically. The most important source of bit-flip errors is found to be induced by the CNOT gate, such that the bit-flip errors induced by idling data qubits are actually negligible with respect to the bit-flips occurring when a CNOT gate is performed. The bit-flip error rate for the CNOT, in the typical range of values for $\kappa_1/\kappa_2$ considered here, is very well approximated by the numerical fit \cite{Chamberland2020_AWS}
\[
p_X^{\text{CX}} = (5.58 \sqrt{\frac{\kappa_1}{\kappa_2}} + 1.68 \frac{\kappa_1}{\kappa_2}) e^{-2\bar n}
\]
where $p_X^{\text{CX}}$ is the sum of all the probabilities of the 12 errors that contain some bit-flip ($X_1$, $X_1X_2$, $X_1Y2$, $X_1Z_2$, $X_2$, etc). A full cycle of QEC requires $2d(d-1)$ CNOT gates. Assuming pessimistically that any single bit-flip error will result in a logical $X_L$ error, and assuming $p_X^{\text{CX}}$ is small, the resulting logical error probability is simply bounded by 
\[
p_{X_L} = 2d(d-1) p_X^{\text{CX}}.
\]} 
\mm{Note that this calculation corresponds to a worst-case scenario, as all such bit-flips do not necessarily lead to a logical failure. For instance, a correlated $X$ error on the control and the target qubit of a CNOT gate of the first round propagates through the second CNOT as a $X_i X_{i+1}$ error on the data, which is a stabilizer and hence does not induce a logical failure.}

\begin{figure}[h]
\includegraphics[width=.5\textwidth]{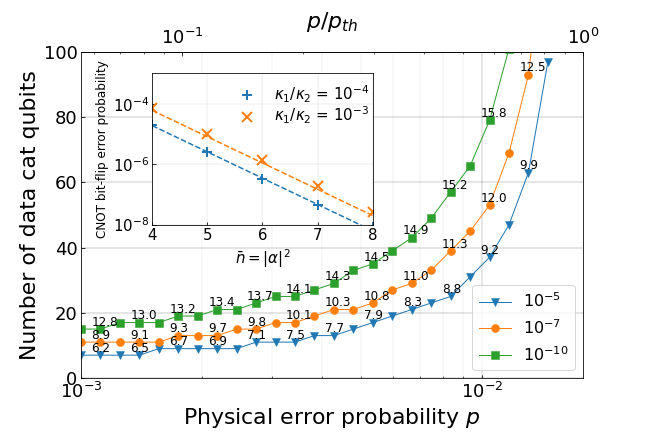}
\caption{
Estimated number of cat qubits per repetition cat qubit used as a quantum memory, versus the physical noise probability $p$, also given in units of the phase-flip threshold value $p_{th}$.
The different plots correspond to different values of the target logical error probability per QEC cycle, and the numbers on the curves correspond to the mean number of photons $\bar n=|\alpha|^2$  in the cat qubits. Inset: Total bit-flip error probability induced by single photon loss at rate $\kappa_1$ during a CNOT gate vs. the mean number of photons $\bar n$. The dotted lines correspond to the numerical fit $p_X^{\text{CX}} = (5.58 \sqrt{\frac{\kappa_1}{\kappa_2}} + 1.68 \frac{\kappa_1}{\kappa_2}) e^{-2\bar n}$ of \cite{Chamberland2020_AWS}.
} 
\label{fig:memory_overhead}
\end{figure}

\mm{This worst-case probability has been taken into account in the estimation of the overall error probability of Figure~\ref{fig:memory_overhead}.}

\begin{figure}[h]
\includegraphics[width=.5\textwidth]{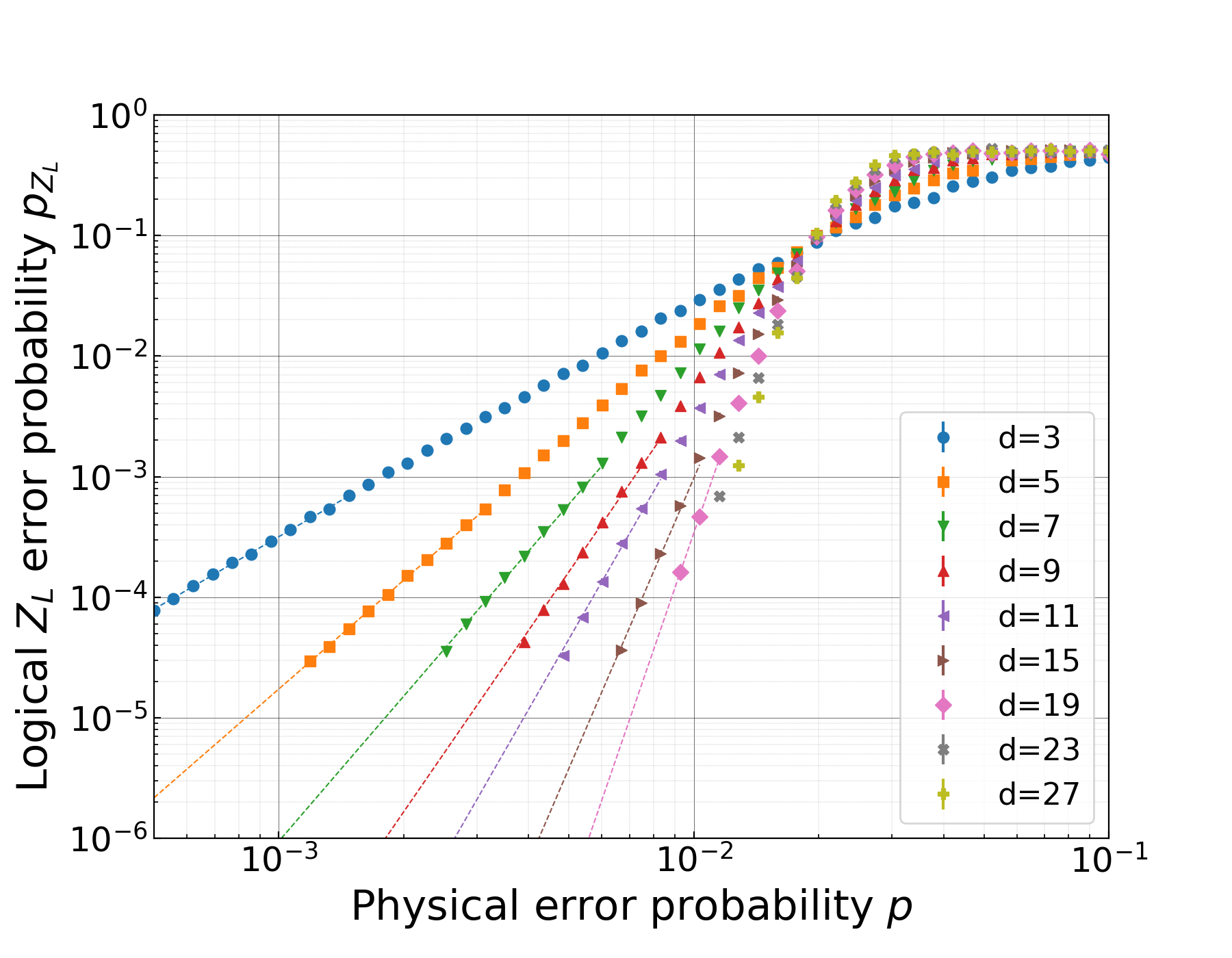}
\caption{
Probability that the error correction circuit of Fig.~\ref{fig:QEC_circuit} induces a logical $Z_L$ error on the repetition cat qubit after the correction is performed. The dotted lines correspond to the asymptotic regime and fit the empirical scaling formula $p_{Z_L} = A(\frac{p}{p_{th}})^{\frac{d+1}{2}}$.
\label{fig:memory_pZL}}
\end{figure}

\paragraph{Logical Z error probability $p_{Z_L}$}
To estimate $p_{Z_L}$, we perform Monte Carlo simulations of the QEC circuit depicted in Fig.~\ref{fig:QEC_circuit} where we neglect physical X errors. Here and in the following simulations, we assume that the classical processing of the measurement outcomes is instantaneous, so no errors are induced on the data qubits while the decoding is performed. In the memory case, we also assume the correction step is perfect, because in this case the correction does not need to be physically applied but rather can be performed in software by updating the Pauli frame~\cite{Knill2005_bis}.

For each run, the repetition cat qubit is initialized in a codeword $\ket{\psi_{in}}_L$. The stabilizers of the code are measured $d$ times as depicted in Fig.~\ref{fig:QEC_circuit}. A last round of perfect stabilizer measurements is performed and the history of measurement outcomes is decoded together with this last perfect measurement outcome. This ensures that, after the perfect correction, the output state $\ket{\psi_{out}}_L$ is back in the code space, either $\ket{\psi_{out}}_L = \ket{\psi_{in}}_L$, in which case the error correction was successful, or a $Z_L$ error occurred, $\ket{\psi_{out}}_L = Z_L \ket{\psi_{in}}_L$. We plot in Fig.~\ref{fig:memory_pZL} the probability $Z_L$ that a logical error occurred for various code distances $d$ and values of the physical noise strength $p$. The phase-flip threshold for this circuit is $p_{\text{th}} = 1.9\%$, which, according to the equation~\eqref{eq:probaOpt}, corresponds to a ratio between the two-photon dissipation rate and a  single photon loss rate $\ktwo/\kone = 220$, close to the value achieved in~\cite{Touzard2018}. For a typical cavity lifetime of 1ms and cat qubits of size \jg{$\bar n = 15$ photons}, this phase-flip threshold of $1.9\%$ corresponds to a CNOT gate time of about \jg{$1.3\mu s$}. It is experimentally reasonable to think that all the other operations can be performed as fast.

Note that the phase-flip threshold for the CNOT error probability is about \jg{$8\%$} (see~\cite{Guillaud2019} and Table \ref{tab:error_models}). For a depolarizing model where idle qubits, state preparation and measurement, and the CNOT gate all fail with probability $p$, and where the CNOT error model is balanced $p_{Z_1} = p_{Z_2} = p_{Z_1Z_2} = p/3$, the fault-tolerance threshold is slightly above $3\%$
~\cite{Suzuki2017}. In our case, a higher gate error probability is tolerated because the phase-flips errors of the CNOT mostly occur on the ancilla cat qubits used for the stabilizer measurement.

\paragraph{Logical error rate and resource overhead}
Combining the logical $X_L$ and $Z_L$ errors, we estimate the minimum number of data cat qubits and the \jg{required average photon number} per cat qubit to achieve a target logical error rate $p_L$ for a quantum memory. In Fig.~\ref{fig:memory_overhead}, we present this physical overhead as a function of the physical error probability $p$.  Quite remarkably, with physical error probabilities of about $1\%$ (corresponding to a CNOT fidelity of $96\%$),  very low logical error probabilities of order $10^{-10}$ per QEC cycle can be achieved for a modest number of 70 modes per repetition cat qubit (twice as much including the ancillary modes) and for experimentally reasonable cat sizes of about 15 photons. Furthermore, with the specific gate realizations of
~\cite{Guillaud2019}, this physical error probability of $1\%$ can for instance be achieved with a two-photon dissipation rate of 125kHz, a cavity mode lifetime of about 1ms and a gate time of about \jg{0.7$\mu$s}. These numbers indicate that using a repetition cat qubit as a quantum memory is a  promising approach to build a  long-lived quantum memory in near term experiments.

\subsection{\label{subsec:transversal operations}Transversal gates}
\begin{figure}[h]
\includegraphics[width=.5\textwidth]{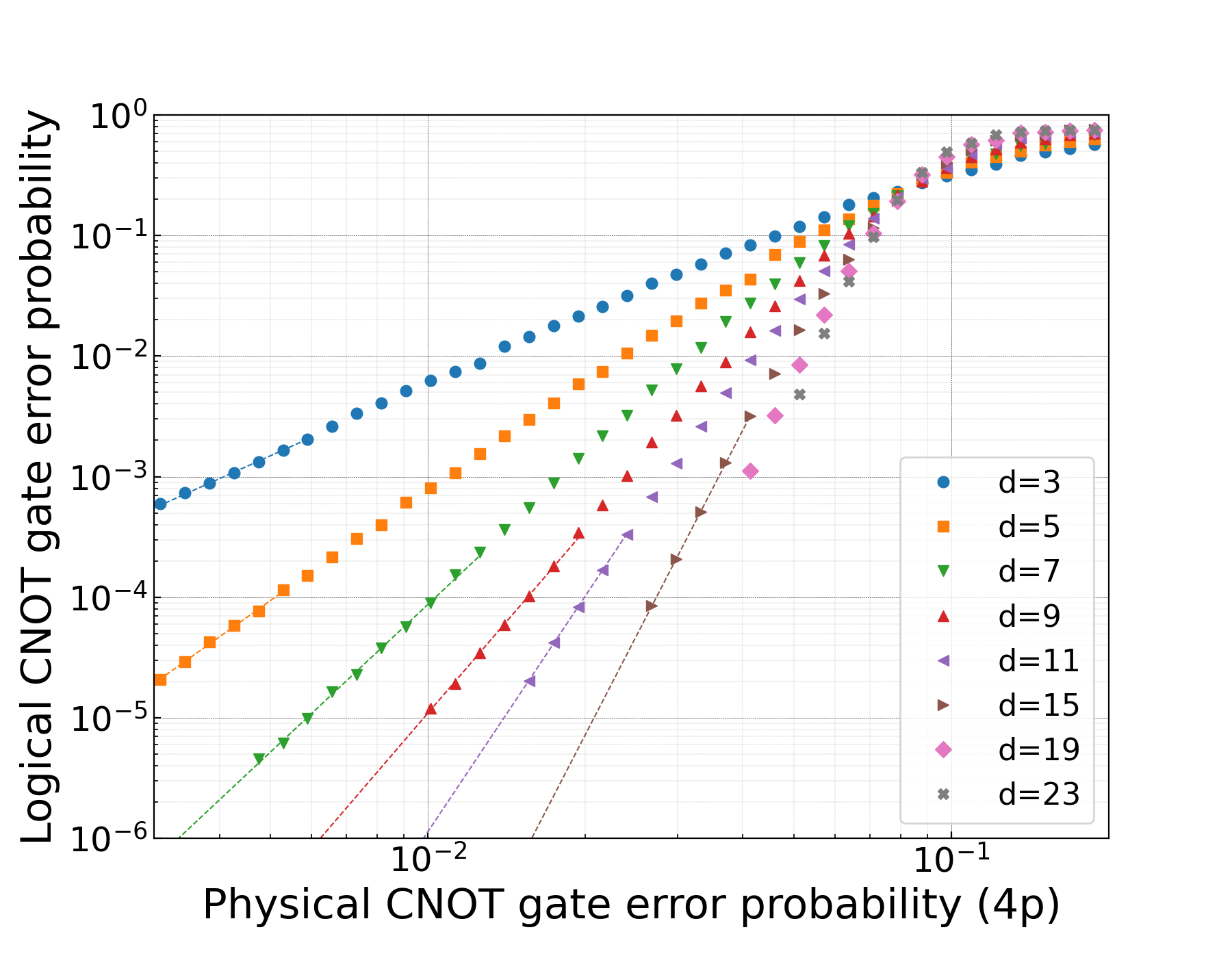}
\caption{Error probability of a transversal logical CNOT gate as a function of the physical CNOT gate error probability given by $4p$ (see~\cite{Guillaud2019}). The asymptotic dotted curves are fits to the empirical scaling formula $A(\frac{p}{p_{th}})^{\frac{d+1}{2}}$. }   
\label{fig:CNOT_performance}
\end{figure}

All logical operations that admit a transversal implementation exhibit a similar performance to the quantum memory. This includes the measurement of $X_L$, the preparation of the logical $\ket\pm_L$ states,  and the logical CNOT gate. The measurement of the $X_L$ operator is done by measuring all the cat qubits in the $X$ basis, followed by a majority vote on the measurement outcomes.
The fault-tolerant preparation of the state $\ket\pm_L$ consists in preparing all the cat qubits in the $\ket +$ state, and performing a full round of error correction as in Fig.~\ref{fig:QEC_circuit}. The phase-flip threshold for this preparation is therefore the same as the quantum memory. The logical CNOT gate is implemented on the codespace by performing a physical CNOT gate between each pair of cat qubits of two different logical codeblocks, followed by a separate round of error correction on each logical block. As it can be seen from Fig.~\ref{fig:CNOT_performance}, the error probability of a logical CNOT gate is similar to that of a quantum memory. 

As shown in~\cite{Guillaud2019}, the set of fault-tolerant gates that can be used to perform universal quantum computation using repetition cat qubits is $\mathcal{S}_L =  \{\mathcal{P}_{\ket\pm_L}, \mathcal{M}_{X_L}, X_L, \text{CNOT}_L, \text{Toffoli}_L \}$. In this section we showed that all these gates except the logical Toffoli can be implemented with very high fidelities for modest code sizes. In the next section, we investigate the performance of this non-Clifford gate.

\section{\label{sec:Toffoli}Non-Clifford operation: the Toffoli gate}

The Eastin-Knill theorem~\cite{Eastin2009} establishes that a set of transversal logical gates cannot be universal for quantum computation. For many quantum codes, the encoded version of the gates in the Clifford group can be implemented transversally on the code, which, by virtue of the Eastin-Knill theorem, prevents non-Clifford gates to be implemented transversally. The fault-tolerant but not transversal construction of a non-Clifford gate is possible but is usually much more expensive in terms of physical resources than the transversal gates. Thus, most of the focus of such constructions has been devoted to find the most efficient strategies to reduce the associated overhead. A long standing leading strategy inspired from gate teleportation techniques~\cite{Gottesman1999} is to prepare encoded versions of magic states and to consume these states as a non-Clifford resource during the computation~\cite{Bravyi2005}. The cost of these techniques, initially very expensive in terms of hardware resources, have been greatly reduced thanks to many years of active research~\cite{Fowler2013,DuclosCianci2013,DuclosCianci2015,OGorman2017,Haah2018,Gidney2019,Litinski2019,Chamberland2020}. In order to avoid magic state distillation and the costly overhead associated with it, a second approach relies on subsystems codes, for which the encoding can be deformed in such a way that the information remains protected from errors but the set of allowed transversal gates changes. This approach includes the ``gauge color codes''~\cite{Bombin2015,Brown2016,Kubica2015bis, Paetznick2013} and other code-switching techniques~\cite{Anderson2014}. Along the same lines, a recent proposal proposed to use code deformation techniques to directly implement a fault-tolerant non-Clifford gate on the surface code~\cite{Brown2020}. A third approach is to combine different codes that have different transversal gate sets using concatenation, to achieve universality in a larger code~\cite{JochymOConnor2014}. Other strategies focus on circuits that are not transversal, yet can still be made fault-tolerant, such as the pieceable fault-tolerant EC where intermediate rounds of error correction are added in well chosen locations of the circuit~\cite{Yoder2016,Takagi2017}, or the flag fault-tolerant EC where additional 
``flag'' ancilla qubits are used to gain more information about the propagating errors~\cite{Chao2018,Chamberland2018}.

In this work, we build on the logical Toffoli gate construction proposed in~\cite{Guillaud2019} and investigate two different strategies for fault-tolerance. The first strategy consists in studying the performance of the circuit depicted in Fig.~\ref{fig:nonFT_ToffoliCircuit}. As explained in~\cite{Guillaud2019}, the intermediate EC in this circuit ensures that the errors do not propagate in an uncontrolled manner. However, in order to ensure the fault-tolerance, one also needs to prevent an accumulation of non-propagating errors in the first and second blocks. As soon as the logical error probability of the Toffoli circuit becomes lower than that of a physical Toffoli gate, the circuit can be made fault-tolerant by using concatenation~\cite{Yoder2016}. The idea of code concatenation is to build a hierarchy of codes within codes iteratively, by replacing all the physical gates in a logical circuit by their logical versions (see e.g~\cite{Nielsen2009}). We argue in Subsection
~\ref{subsec:non-FT Toffoli} that with experimentally reasonable physical error rates, the logical error rate of the Toffoli circuits is well below that of the physical one; thus making it possible to use code concatenation.

In Subsection~\ref{subsec:FT Toffoli}, we study a second strategy that avoids concatenation and achieves a higher phase-flip threshold and an improved scaling, but comes at the expense of a more complex error correction circuit based on three ingredients. First, the accumulation of non-propagating errors is prevented using the pieceable fault-tolerant protocol described in ~\cite{Yoder2016, Guillaud2019}. Second, this pieceable fault-tolerant protocol requires the measurement of the stabilizers of the code in the middle of the logical Toffoli circuit, at a point where these stabilizers are no longer Pauli operators, but have evolved to Clifford operators under the action of the non-Clifford pieces of the Toffoli circuit. Last, we use a Steane-style error detection ~\cite{Steane1998} decoded with a majority vote on the target block instead of the usual stabilizer measurements decoded with the MWPM decoder used everywhere else in this work.

\begin{figure}[h]
\includegraphics[width=.2\textwidth]{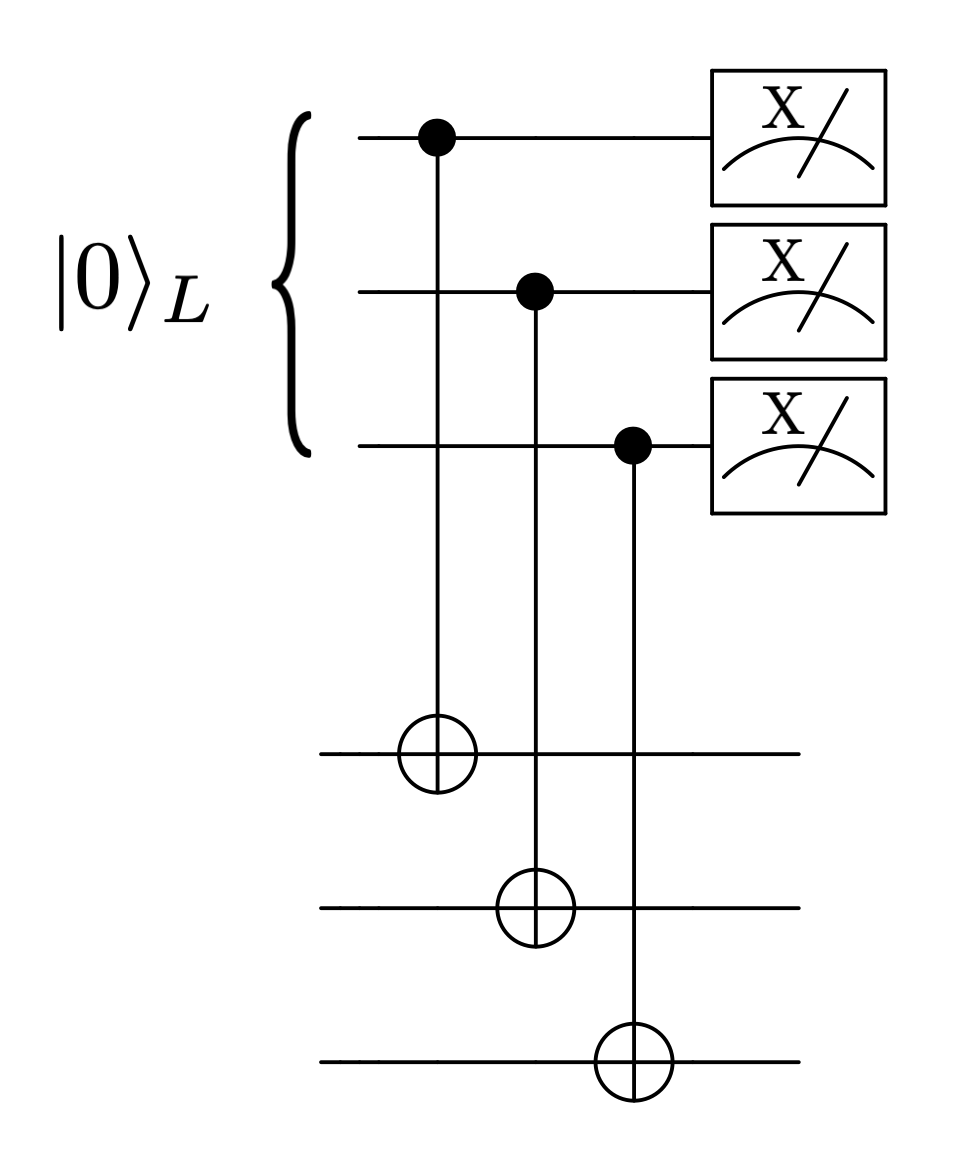}
\caption{Steane error correction for a distance-3 repetition code protecting against phase-flips. The protocol requires the preparation of the $\ket 0_L$ state on a logical ancillary qubit and the phase-flip errors of the logical data qubit (bottom three lines) are copied on the ancillary block by the CNOT gates and detected by the ancilla measurements.} \label{fig:Steane_EC}
\end{figure}

For the sake of completeness, we briefly recall the Steane QEC procedure. A logical ancilla qubit is prepared in the state $\ket 0 _L$, and logical CNOT is applied between the logical ancilla qubit as control and the logical data as target, as depicted in Fig.\ref{fig:Steane_EC}. Since the control is prepared in the $\ket 0 _L$, the logical CNOT has no effect on the logical state, but the phase-flip errors of the target are copied on the ancilla block and then detected through a simple majority vote. Note that in the usual case, bit-flip errors can propagate from the ancilla block to the target block via the CNOT, which usually requires the logical ancilla state to be verified before it can be used in this protocol. Here, because the cat qubits have no bit-flip errors, and because the CNOT gates are bias-preserving, this verification step is unnecessary. \jg{The fault-tolerant preparation of the $\ket{0}_L$ can be achieved as in \cite{Chamberland2020_AWS}. We note that this preparation is non deterministic, which implies an extra overhead. While we have taken into account the stochastic errors introduced by the preparation step given in Table \ref{tab:error_models}, the overhead associated with the synchronization of the preparation of the $\ket{0}_L$ with the pieces of the circuit is not included in our study.}

\subsection{\label{subsec:non-FT Toffoli}Fault-tolerance with concatenation}

\begin{figure}[h]
\includegraphics[width=.5\textwidth]{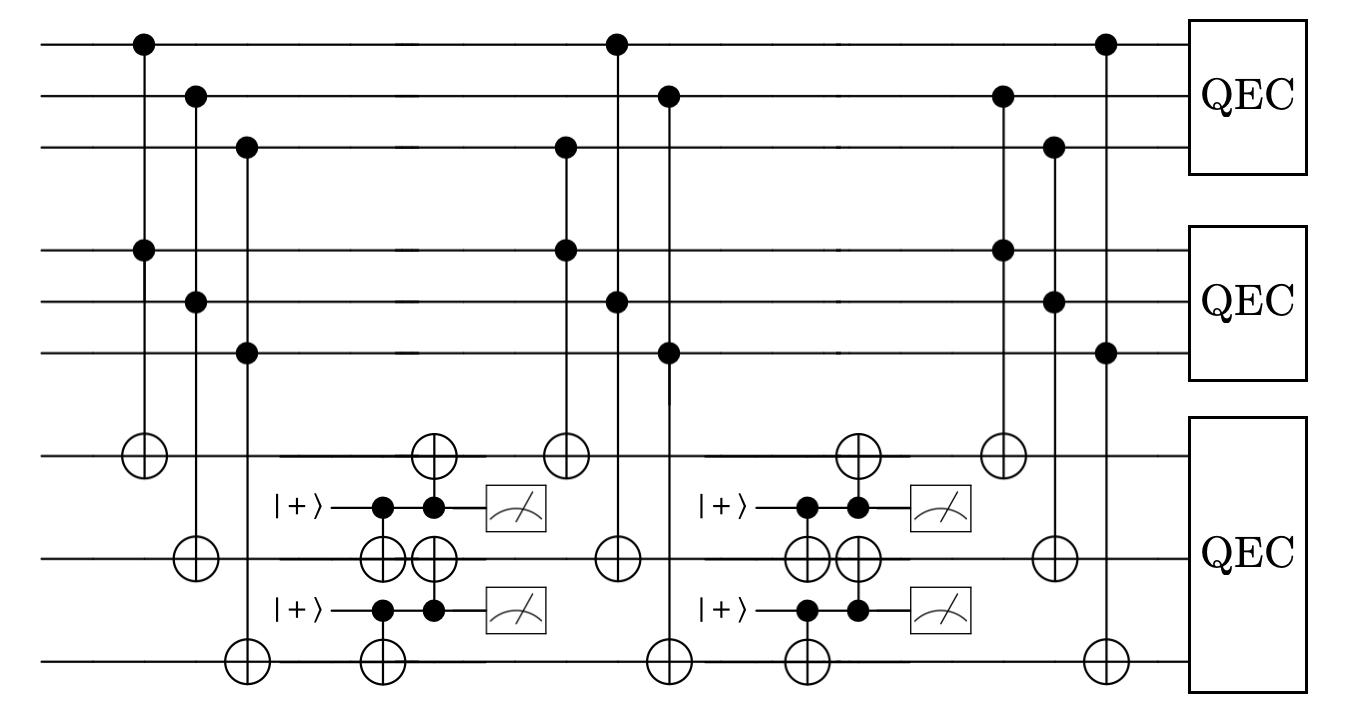}
\caption{
Logical Toffoli circuit for distance 3 repetition cat qubits. After each round of transversal Toffoli gates, a single round of stabilizer measurement is performed on the target block. The outcome of the measurement is decoded together with the history of all outcomes and an appropriate correction is applied. After the circuit, a full error correction stage is performed on all three blocks.
} 
\label{fig:nonFT_ToffoliCircuit}
\end{figure}

In this subsection, we numerically simulate the circuit of Fig.~\ref{fig:nonFT_ToffoliCircuit} to estimate the value of the phase-flip threshold with concatenation. This corresponds to \jg{the so-called \textit{breakeven}, where} the value of the physical error probability below which the logical Toffoli error probability is smaller than the physical one. As soon as the phase-flip error probability is below this threshold, the repetition code can be concatenated with a second repetition code. Note that the concatenation we consider here is that of two repetition codes, in order to make the logical Toffoli circuit fault-tolerant with respect to phase-flip errors only, and to get a phase-flip threshold. This is different from the concatenation that we discussed in the introduction, where the higher level code is chosen to be a code possessing an accuracy threshold with respect to both phase-flip and bit flip errors. 

The logical Toffoli gate is built using $d^2$ physical Toffoli gates through the round-robin construction
\[
\prod \limits_{i,j \in [\![ 0, d - 1 ]\!]} \text{CCX}(i,j,k(i,j))
\]
where $\text{CCX}(i,j,k)$ denotes a physical Toffoli gate between the $i$-th qubit of the first control block, the $j$-th qubit of the second control block, and the $k$-th qubit of target block. Note that $k(i,j)$ can actually be any mapping $[\![ 0, d-1 ]\!] \times [\![ 0, d - 1 ]\!] \rightarrow [\![ 0, d - 1]\!]$, since the gates $\text{CCX}(i,j,k_1)$ and $\text{CCX}(i,j,k_2)$ act identically on the codespaces of the three logical qubits, and we set $k(i,j) = j$ for the rest of the paper.

There are two reasons this construction is not fault-tolerant as such. First, because this circuit is not transversal and any qubit of the target block is connected to \textit{all} the qubits of the first control block, a single Z error acting on a qubit of the target block can be copied many times on the first control block, possibly leading to a logical failure. For example, a Z error occurring on a qubit of the target block before the circuit is executed propagates to the same Z error on the target block, plus a logical CZ gate between the two logical control qubits. This first problem can be solved following the pieceable fault-tolerant method of ~\cite{Yoder2016}. More precisely, we split the circuit containing $d^2$ physical Toffoli gates into $d$ transversal pieces of $d$ Toffoli gates each. As shown in Fig.~\ref{fig:nonFT_ToffoliCircuit}, between two pieces, a round of error correction is performed on the target block to catch errors before they spread to the control blocks. Importantly, because the target X operator commutes with the CCX gate, the stabilizers of the target block $\{ X_iX_{i+1}, i \in [\![ 0, d-1 ]\!]\}$ are left unchanged by the CCX gates of the circuit and can be measured at any point in the circuit with the circuit of Fig.~\ref{fig:stabilizer_measurement_circuit}. The logical Toffoli circuit is executed as follows: after each of the first $d-1$ transversal pieces of CCX gates, a single round of stabilizer measurement is performed on the target block. After each of these pieces, say the $k$-th one, the $k$ outcomes from all the previous measurements rounds are decoded together using a minimum weight matching decoder. The corresponding correction is applied before the $(k+1)$-th piece of the circuit is executed. After the last piece is executed, the usual error correction, composed of d rounds of stabilizer measurements and correction, is performed on the three codeblocks. The fact that a single round of stabilizer measurement is enough during the intermediate error correcting steps can be intriguing. Indeed, since the measurements themselves are faulty, they usually need to be repeated a certain number of times, that scales linearly with the code distance, before the outcome of the first measurement, decoded together with the ones following, can be trusted. Here, a single round is executed independent of the code size, but is decoded using the full history of the previous measurement outcomes. The history of the $Z$ correction applied on the target block is also kept in memory. Thus, after all the $d$ pieces have been executed, a final decoding on all $d$ syndrome outcomes is performed and it becomes possible to know \textit{a posteriori} which target Z errors have propagated to CZ errors between control qubits and to correct the corresponding CZ before the final QEC round on the control blocks. This is possible because the Z corrections performed on the target block anti-commute with the constant stabilizers of the target block, thus travel without being projected and can be undone later if needed.

\begin{figure}[h]
\includegraphics[width=.5\textwidth]{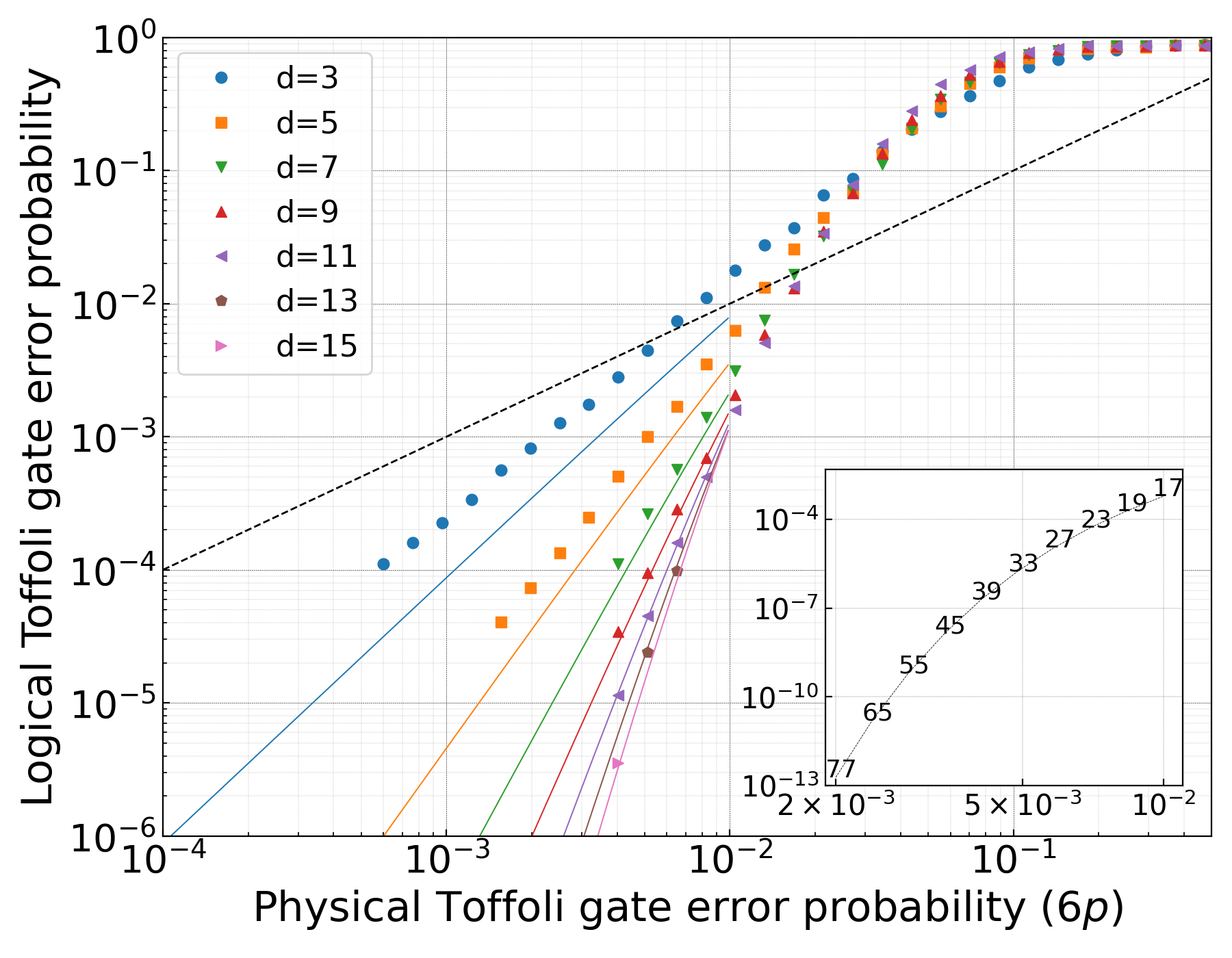}
\caption{Error probability of the logical Toffoli gate implemented by the circuit of Fig.~\ref{fig:nonFT_ToffoliCircuit} as a function of the  error probability of  physical cat qubit Toffoli gates. The colored cross are numerical simulation results, the plain lines are computed analytically and correspond to the error probability of this circuit in the case where the final QEC stage is perfect (see main text). The dotted black line is the identity and serves as a guide to the eye to visualize the ``break-even point'' below which the error probability of the logical Toffoli circuit is smaller than that of the physical Toffoli gate. Inset: Minimal value of the logical Toffoli error probability achievable for a given physical error probability. The optimal distance realizing this minimum is indicated on the curve.}
\label{fig:nonFT_ToffoliPerformance}
\end{figure}

The second reason the circuit is not fault-tolerant is because of the accumulation of non-propagating errors on the control qubits. Indeed, each qubit of the two control blocks undergo $d$ gates without the stabilizers of these qubits being measured. Therefore, without further considerations, when increasing the code distance $d$, the probability of Z errors on these control blocks increases and eventually exceeds the fault-tolerance threshold of the repetition code. One way to handle this problem is by concatenation with another repetition code. Indeed, we will see throughout the rest of this subsection that despite the accumulation of non-propagating errors, the circuit of Fig.~\ref{fig:nonFT_ToffoliCircuit} yields very low logical error probabilities. The existence of a reasonable break-even point (a physical error probability for which one can find a code distance $d$ yielding a lower logical error probability) proves the existence of a concatenation phase-flip threshold. Concatenating a distance d repetition code with itself produces a repetition code of distance $d^2$. The circuit implementing a logical Toffoli on the concatenation of these two codes is very similar to the one depicted in Fig. \ref{fig:nonFT_ToffoliCircuit}, except that it now includes error correcting steps on the control blocks every $d$ steps. The distance can be further increased by raising the number of levels of concatenation (the concatenation of $k$ repetition code produces a distance $d^k$ repetition code), while the number of steps between two rounds of error correction remains constant (equal to $d$).

In Fig.~\ref{fig:nonFT_ToffoliPerformance}, we simulate the circuit of Fig.~\ref{fig:nonFT_ToffoliCircuit} taking into account a circuit-based error model with error models provided in Table~\ref{tab:error_models}. As mentioned above, the absence of a phase-flip threshold can be explained by the accumulation of non-propagating errors: for all values of the physical error probability $p$, there is a finite optimal value of the code distance that achieves a minimum logical error probability. \jg{However, even though this circuit does not possess a fault-tolerance threshold, it can still be used to produce a logical Toffoli gate whose failure probability is well below that of the physical Toffoli gate. In order to emphasize this, we also plot in Fig.~\ref{fig:nonFT_ToffoliPerformance} the identity line to visualize the ``break-even'' point below which the fidelity of the logical Toffoli gate obtained with circuit Fig.~\ref{fig:nonFT_ToffoliCircuit} becomes higher than the fidelity of the bare Toffoli gate. In this regime, code concatenation becomes possible and allows to get a phase-flip threshold}. Thus, the phase-flip threshold for this circuit used with concatenation is slightly below $2\%$.

In the case where $p$ is small and for large enough code distance $d$, the infidelity of the circuit is dominated by the accumulation of non-propagating errors on the control blocks. The probability that the circuit fails due to the errors in the control blocks exceeds by several orders of magnitude both the failure probability due to the target block errors or the failure probability of error correction blocks. A good estimate of this optimal code distance can be obtained by assuming the QEC steps are perfect. In this case, the probability of a logical $Z_L$ error on either of the control blocks is simply given by the probability of accumulating more than $\lfloor d/2 \rfloor$ errors
\[
p_{Z_L} = \sum \limits_{k = \lfloor d/2 \rfloor+1}^d \binom{d}{k}p'^k(1-p')^{d-k}
\]
where $p' \approx dp$ is the probability that a given physical qubit of a logical control block is corrupted by a $Z$ error during the circuits execution (the approximation $p' \approx dp$ is valid as far as $dp\ll 1$). This infidelity is plotted in plain lines, and, as expected, fits well the numerical values in the regime where the physical error probability $p$ is small and the code distance $d$ is large, for which the logical error is entirely set by the accumulation of non-propagating errors.

We used this asymptotic formula to estimate the optimal code distance $d$ and the associated logical error probability for a given physical Toffoli error probability, as it is precisely the region of the curves where the formula fits very well the numerical values. The results are plotted in the inset of Fig.~\ref{fig:nonFT_ToffoliPerformance}. As it can be observed, even without concatenation, a physical error probability of about .25\% per Toffoli gate on cat qubits yields logical error rates of about $10^{-10}$ with as few as 60 modes. Now, if the physical error probability per cat qubit Toffoli gate is about $1\%$, the same logical error probability of $10^{-10}$ can be achieved with one level of concatenation, concatenating a 9 mode repetition code with a 60 mode one. According to the error model for the Toffoli implementation of~\cite{Guillaud2019}, and assuming a two-photon dissipation rate of $\ktwo/2\pi=1$MHz, this physical error probability can be achieved for a cavity lifetime of 1ms. With the recent progress in 3D superconducting cavities, this long lifetime can be typically achieved with cylindrical postcavities~\cite{Reagor2016}.

\subsection{\label{subsec:FT Toffoli}Fault-tolerance without concatenation}

For the circuit of Fig.\ref{fig:nonFT_ToffoliCircuit} to exhibit a phase-flip threshold without any concatenation, it is necessary to place additional rounds of error correction on the control blocks in such a way that the number of time-steps between two rounds of error detection does not increase with the code distance. Ideally, we would like to perform a round of error correction on all three logical blocks after each of the transversal pieces of the circuit. This task is complicated by the fact that the stabilizers of the control blocks are not constant throughout the circuit. We label the three logical qubits $A$, $B$ and $C$, where $C$ is the logical target block and denote by $X^A_i$ the $X$ Pauli operator acting on the $i$-th physical qubit of block $A$, where all subscripts are taken modulo the code distance $d$. The $k$-th piece of the circuit $P_k$ consists of $d$ transversal Toffoli gates, where the control qubits of block $A$ have been shifted by $k-1$
\[
P_k = \prod \limits_{i = 0}^{d-1} \text{CCX}(i-k+1,i,i).
\]

\begin{figure}[h]
\includegraphics[width=.35\textwidth]{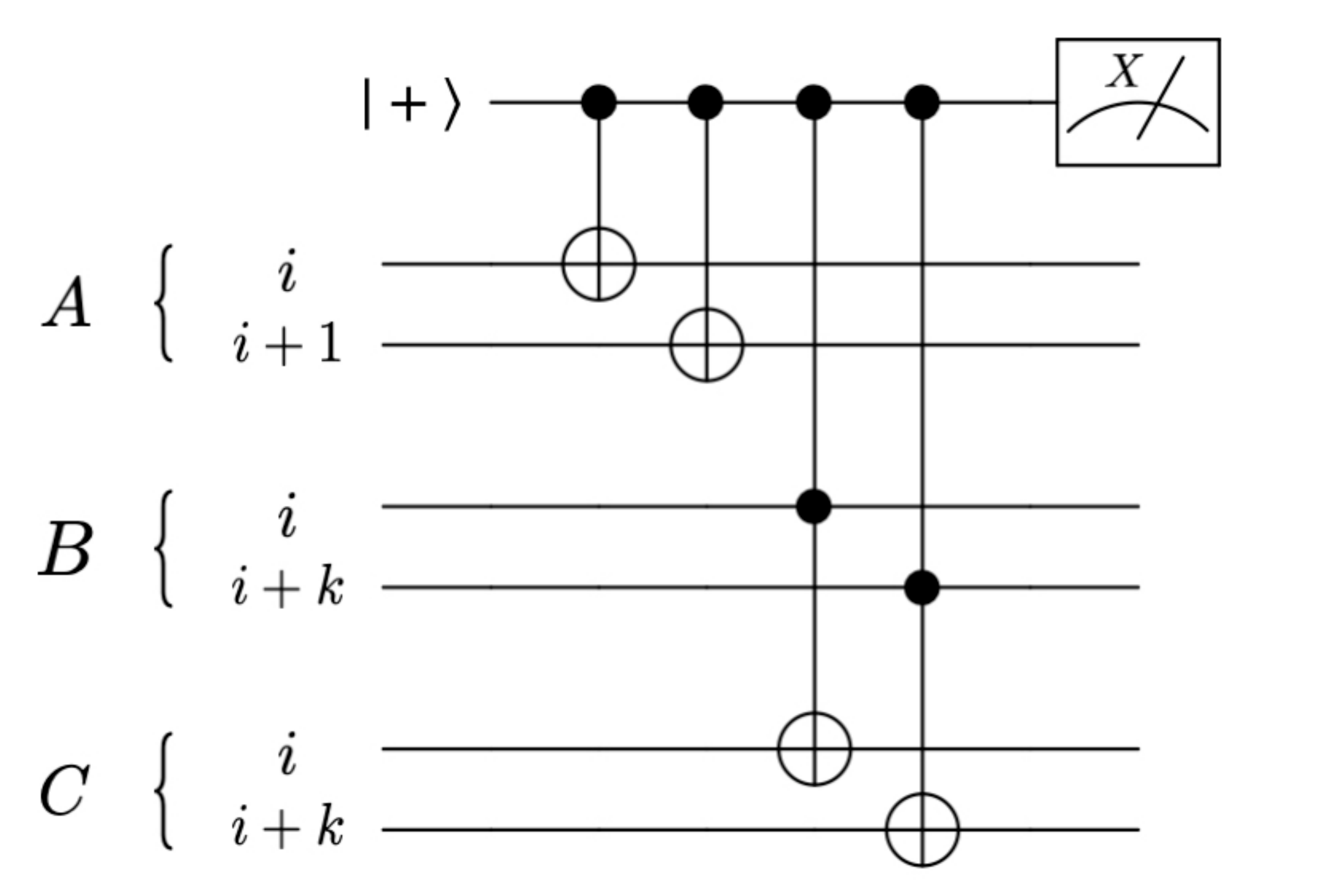}
\caption{Measurement circuit of the Clifford stabilizer $X^A_i X^A_{i+1} \text{CX}^{B,C}(i,i)\text{CX}^{B,C}(i+k,i+k)$.
} 
\label{fig:CliffordStabilizerMeasurement}
\end{figure}

Let us have a look at the value of the non-constant stabilizers of the two control code blocks $X^O_iX^O_{i+1}$, $O \in \{A, B\}$, $i \in [\![ 0, d - 1 ]\!]$, after $k$ pieces of the circuit have been executed. Noting $U_k = \prod \limits_{j \in [\![ 1, k]\!]} P_{j}$, the stabilizers of the two controls blocks $A$ and $B$ become under conjugation by this unitary

\begin{align*}
&S_{i,k}^A:=U_k X^A_i X^A_{i+1} U_k^\dagger\\ 
&\qquad= X^A_i X^A_{i+1} \text{CX}^{B,C}(i,i)\text{CX}^{B,C}(i+k,i+k) \\
&S_{i,k}^B:=U_k X^B_i  X^B_{i+1} U_k^\dagger \\
&= X^B_i X^B_{i+1} \times \prod \limits_{j = 0}^{k-1} \text{CX}^{A,C}(i-j,i)\text{CX}^{A,C}(i+1-j,i+1)
\end{align*}

where $\text{CX}^{R,S}(i,j)$ denotes the CX gate between the $i$-th qubit of block $R$ acting as the control and the $j$-th qubit of block $S$ acting as the target. Note that the stabilizers of the control block $C$ are constant throughout this evolution
$$
S_{i,k}^C:=U_k X^C_i X^C_{i+1} U_k^\dagger = X^C_i X^C_{i+1}.
$$
The evolution of the stabilizers of the control blocks leads to a few issues that need to be handled carefully, to ensure the existence of a phase-flip threshold. 

\begin{figure*}[t!]
\includegraphics[width=\textwidth]{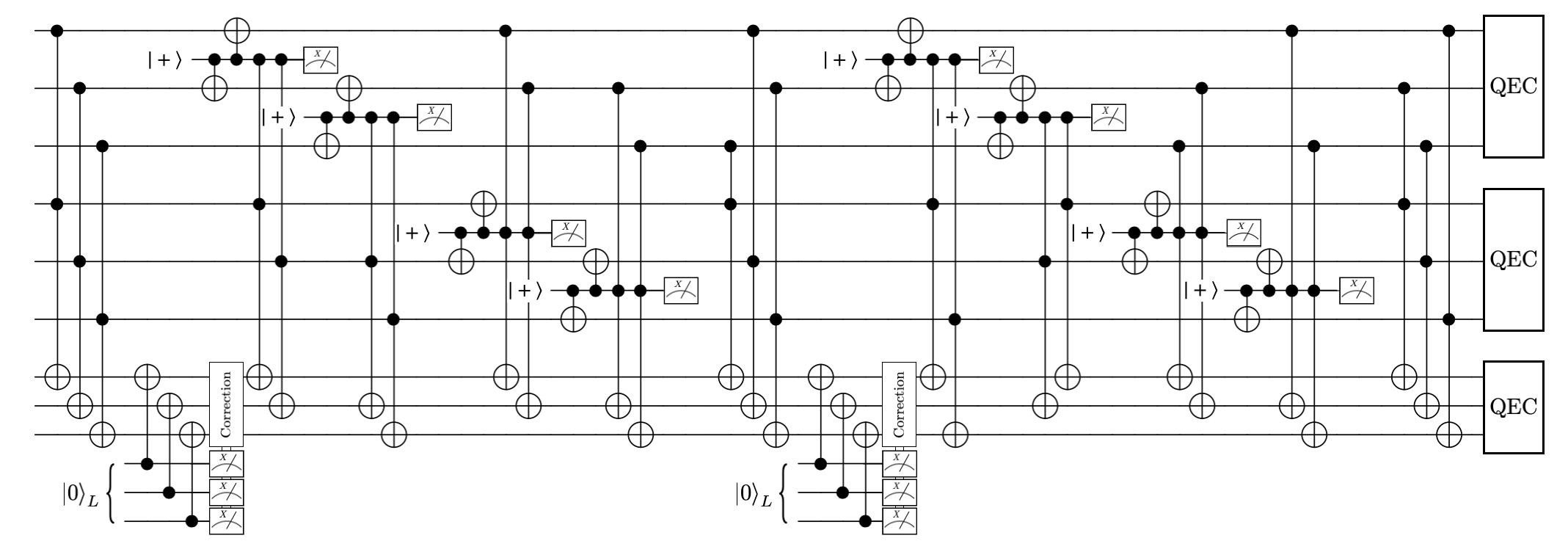}
\caption{A fault-tolerant Toffoli circuit without concatenation. After each of the first $d-1$ pieces of the circuit (here, $d=3$), a round of Steane error correction is performed on the target block, followed by the measurement of the Clifford stabilizers on the control blocks.
\label{fig:FT_ToffoliCircuit}}
\end{figure*}

First, the unitary $U_k$ does not belong to the Clifford group, but to the third level of the Clifford hierarchy~\cite{Gottesman1999}. It maps the Pauli stabilizers of the control blocks to Clifford operators. Nevertheless, these Clifford stabilizers can be measured using CCX gates and Clifford gates. The stabilizer $S_{i,k}^A$ can be measured in the standard way using one ancilla qubit with the circuit depicted in Fig.~\ref{fig:CliffordStabilizerMeasurement}. Importantly, the measurement of the non-constant stabilizers is bias-preserving as the CX and CCX gates possess this property. 

Second, the weight of the stabilizers of control block $A$, $S_{i,k}^A$, is constant at all intermediate steps of the circuit (here, by the weight, we mean the number of physical qubits in the support of the associated observable). Unfortunately, this is not the case for the stabilizers of control block $B$, $S_{i,k}^B$, whose weights grow linearly with $k$ the number of pieces. The asymmetry between the two control blocks is a consequence of the particular choice of ordering for the physical Toffoli gates. A symmetric ordering causes the weight of the stabilizers of both logical blocks to grow linearly with the code distance, but unfortunately it is not possible to order the gates such that the weights of all stabilizers be bounded by a constant. This implies that an increasing depth-$k$ circuit might be needed to measure these stabilizers in the same fashion as  in Figure~\ref{fig:CliffordStabilizerMeasurement}. This scaling of the measurement time with the code distance $d$ prevents the existence of a phase-flip threshold.

The solution that we propose to get around this problem is to measure a different set of Clifford observables of constant weight instead of the stabilizers of block $B$. We chose these observables in such a way that the action of the circuit on the codespace is not modified by their measurement, while their measurement still reveals the value of the actual stabilizers. We call these Clifford observables, the `` modified stabilizers''. One further trick here is to first perform a round of error correction on the target block $C$ before measuring the non-constant stabilizers of block $A$ and the modified ``$B$-stabilizers''. Let us first assume that we can perform an ideal (fault-less) error correction on the target register, mapping the state of the target block $C$ back to the code space. This means that $X^C_i X^C_{i+1} = +1$ for all $i$. Note that 

\begin{multline*}
\prod \limits_{j = 0}^{k-1} \text{CX}^{A,C}(i-j,i)\text{CX}^{A,C}(i+1-j,i+1) = \\ \text{CX}^{A,C}(i+1-k,i)\text{CX}^{A,C}(i+1,i+1) \\
\times \prod \limits_{j = 1}^{k-1} \text{CX}^{A,C}(i+1-j,i)\text{CX}^{A,C}(i+1-j,i+1)
\end{multline*}
and that 
\begin{multline*}
    \text{CX}^{A,C}(i+1-j,i)\text{CX}^{A,C}(i+1-j,i+1)=\\\frac{1}{2}(I+Z^A_{i+1-j})+\frac{1}{2}(I-Z^A_{i+1-j})X^C_{i}X^C_{i+1}.
\end{multline*}
As $X^C_i X^C_{i+1} = +1$, we have
$$
S_{i,k}^B=X^B_i X^B_{i+1}\text{CX}^{A,C}(i+1-k,i)\text{CX}^{A,C}(i+1,i+1),
$$
which admits a constant weight now. It is important to note that these constant-weight ``modified $B$-stabilizers'' only commute with the $A$-stabilizers if the state of the block $C$ is in the codespace.

The remaining question is whether this procedure still works when the error correction step on the target block is imperfect, thus mapping imperfectly the state of the logical block $C$ to the codespace. In this case, the set of ``modified $B$-stabilizers'' may not commute with the $A$-stabilizers, thereby forbidding a simultaneous measurement of these two sets. Indeed, in the current error correction approach, the imperfection of the $C$-stabilizer measurements is compensated by the repetition of these measurements and a MWPM decoder. This procedure however requires to repeat the measurements  a number of times that scales linearly with $d$ and during which we cannot measure the $A$ and $B$ stabilizers. The final trick to get around this issue is to replace the current error correction procedure of the target block by a single round of Steane-style error correction. Indeed, while the Steane-style error correction step can still be faulty, the output errors are not correlated to the input errors. This means that the measurements of the subsequent $A$ and $B$ stabilizers might be faulty, but these errors remain independent and therefore one can still hope to achieve a phase-flip threshold. The full circuit for the logical Toffoli gate, including the different error correction steps, is depicted in Fig.~\ref{fig:FT_ToffoliCircuit}.

\begin{figure}[h]
\includegraphics[width=.5\textwidth]{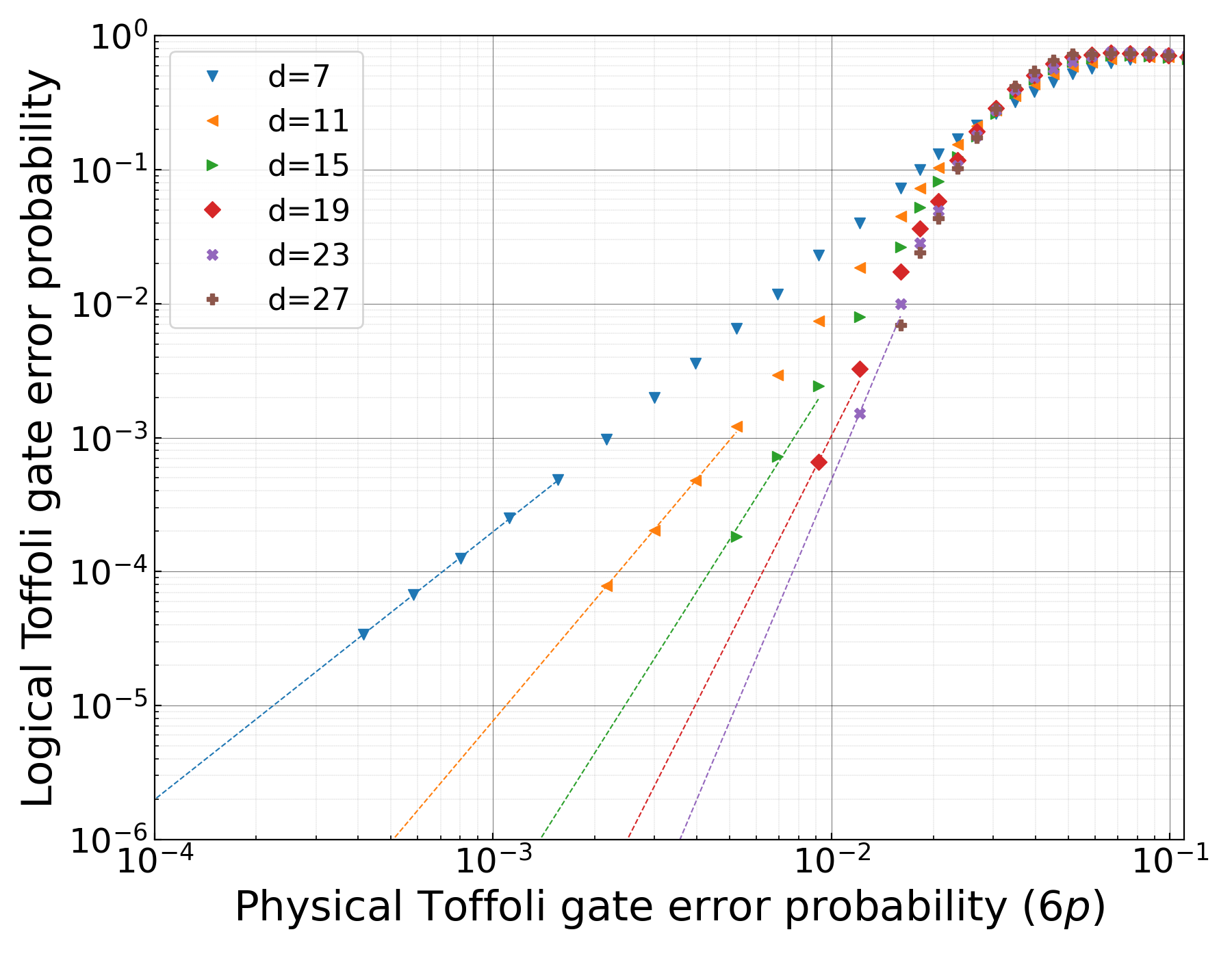}
\caption{Monte Carlo simulations of the circuit of~Fig.~\ref{fig:FT_ToffoliCircuit} using a circuit-based error model. Here, we plot the error probability of the logical Toffoli gate as a function of the error probability of the physical  Toffoli gate. In the asymptotic regime where the physical error probability is small, the logical error probability now with $\lceil\frac d 4\rceil$ instead of the usual $\lceil\frac d 2\rceil$ that we get in the memory case. This is a consequence of the fact that the errors of a single physical qubit of the target block can spread to two different physical qubits within the same logical control block through the stabilizer measurements. Here, the curves are fit to the empirical scaling formula $A(\frac{p}{p_{th}})^{\frac{d+1}{4}}$. }
\label{fig:FT_ToffoliPerformance}
\end{figure}

We perform the Monte-Carlo simulations of this circuit, using a circuit-based error model including the error models provided in the Table~\ref{tab:error_models}. The simulation results are plotted in Fig.~\ref{fig:FT_ToffoliPerformance}. These simulations indicate the existence of a threshold corresponding to a physical Toffoli error probability sightly below $3\%$. A typical physical error probability of $1\%$ that can be achieved with the parameters of the previous subsection should result in a logical Toffoli error probability of $10^{-10}$ with as few as 90 data modes. This important overhead reduction, with respect to the previous concatenated case, comes at the expense of a fault-tolerant preparation of logical $\ket{0}_L$ states that will be consumed by the Steane EC protocol. This logical preparation can be performed by initializing each mode in the coherent state $\ket{\alpha}$ followed by $d$ rounds of $X_iX_{i+1}$ parity measurements and correction by MWPM. This requires to allocate a memory register in which the states are constantly prepared, maintained by EC, and consumed by logical Toffoli gates when needed.

\section{\label{sec:architecture}Towards a practical architecture}
We have investigated the error rates and resource overhead that could be expected using repetition cat qubits. In this work, we have not yet considered the physical restrictions imposed by the particular experimental implementation of the scheme. Typically, a realistic scheme for large scale fault-tolerant quantum computation should possess the following features: a high accuracy threshold such that error rates well below this value can be achieved in the experiments, a universal set of logical gates that can be implemented with a reasonable resource overhead, and an architecture that can be scaled up to a size where the logical error rates match those needed for the targeted computation. The first and the last points are the strongest assets of the surface code approach, and the main reason for its popularity. Indeed, this code combines the advantage of a high accuracy threshold around $1\%$ for a depolarizing noise model~\cite{Raussendorf2007} and a 2D spatial arrangement of the physical qubits requiring only low-weight stabilizer measurements between nearest neighbours. Here, the transversal Clifford operations presented in Section~\ref{sec:CliffordOperations} are compatible with a 2D architecture and using only couplings between neighbouring qubits. However, the two circuits proposed in Section~\ref{sec:Toffoli} exploit an all-to-all coupling between the data cat qubits of the two logical control blocks. Within the particular circuit QED framework that we have in mind for the experimental implementation of repetition cat qubits, this kind of connectivity is less practical and rises a major challenge. Yet, it is worth noting that we anticipate very low logical error rates with only a few tens of cat qubits per logical qubit, which is a drastically lower overhead than those usually envisioned in other QEC schemes. Therefore, the general constraints on the connectivity graph of the physical qubits may be easier to satisfy for near term experiments involving a small number of cat qubits, yet achieving low logical error rates.

\begin{figure}[t!]
\includegraphics[width=.5\textwidth]{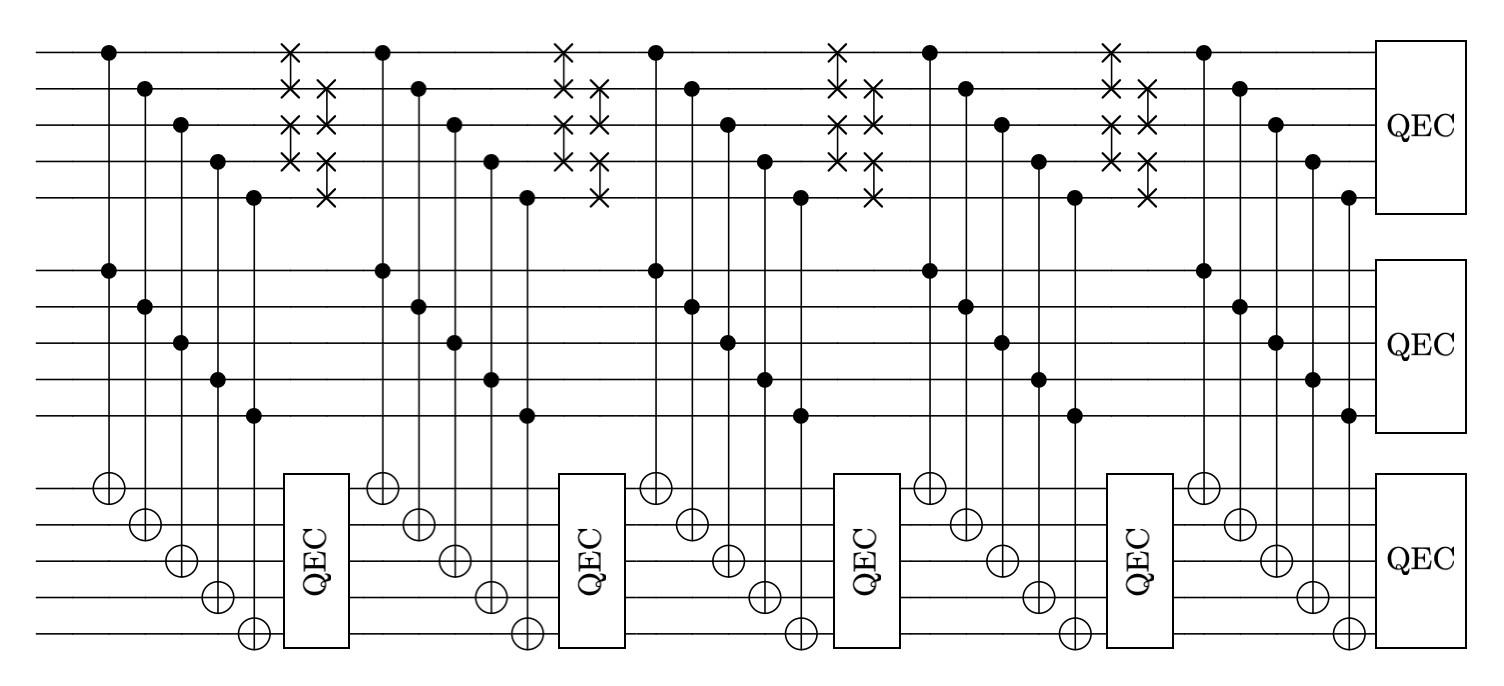}
\caption{Logical Toffoli circuit for distance 5 repetition cat qubits including physical SWAPs on the first control block during error correcting stage on the target. The physical SWAP gates ensure a great simplification of the connectivity graph for the implementation of the logical Toffoli gate.}
\label{fig:nonFT_ToffoliCircuit_SWAP}
\end{figure}

While the optimal layout of a large scale quantum computer based on repetition cat qubits is not known yet, here we provide a few possible directions. The connectivity graph for the logical Toffoli circuit of Section~\ref{subsec:non-FT Toffoli} can be made ``local'' by swapping the data qubits of the first control block appropriately, as depicted in Fig.\ref{fig:nonFT_ToffoliCircuit_SWAP}. Each physical cat qubit of a given repetition cat qubit now only needs to be coupled to a single cat qubit of another repetition cat qubit. Yet, the intermediate rounds of error correction on the target block are still needed to prevent the propagation of errors. The particular ordering of the physical Toffoli gates in Fig~\ref{fig:nonFT_ToffoliCircuit_SWAP} differs from the circular permutations previously considered. This particular choice corresponds to a permutation that can be implemented with parallel SWAPs in two steps, independently of the code distance. One may wonder whether the same trick can be applied to the second Toffoli circuit (Fig.~\ref{fig:FT_ToffoliCircuit}) or a similar fault-tolerant circuit. The existence of a Toffoli circuit ordering that allows us both to measure constant weight Clifford operators for the intermediate error correction steps on the logical control blocks and that can be implemented using a constant depth circuit of SWAP gates is an open problem and requires further investigation. 

An important question is whether the SWAP operations can be performed in a bias-preserving manner. The answer is yes, since a SWAP gate can be implemented using three CNOT gates, but there is a more direct way to implement a bias-preserving SWAP gate between two physical cat qubits. In the same spirit as the CNOT gate, this is done by replacing the regular two-photon dissipators $\hat{L}_{\ha} = \ha^{2} - \alpha^2$ and $\hat{L}_{\hb} = \hb^{2} - \alpha^2$ by the following time-dependent operators that combine both modes
\begin{align*}
    \hat{L}_{\ha} (t) &= \ha^{2} - \tfrac 12 \ha\hb(1 - e^{2i\frac{\pi}{T}t}) - \tfrac 12 \alpha^2(1 + e^{2i\frac{\pi}{T}t}), \\
    \hat{L}_{\hb} (t) &= \hb^{2} - \tfrac 12 \ha\hb(1 - e^{-2i\frac{\pi}{T}t}) - \tfrac 12 \alpha^2(1 + e^{-2i\frac{\pi}{T}t}).
\end{align*}
where $t \in [0, T]$ and $T$ is the SWAP gate time. The instantaneous joint kernel of these operators is the four dimensional Hilbert space spanned by the coherent states 
$$\ket{\alpha, \alpha}, \ket{-\alpha, -\alpha}, \ket{\alpha e^{i\frac{\pi}{T}t}, -\alpha e^{-i\frac{\pi}{T}t}}, \ket{-\alpha e^{i\frac{\pi}{T}t}, \alpha e^{-i\frac{\pi}{T}t}}.$$
Recalling that $\ket{0} \approx \ket{\alpha}$ and $\ket{1} \approx \ket{-\alpha}$, these two dissipation channels implement the correct mapping corresponding to a SWAP gate:
\begin{align*}
    \ket{\alpha,\alpha} &\rightarrow \ket{\alpha,\alpha}\\
    \ket{-\alpha,-\alpha} &\rightarrow \ket{-\alpha,-\alpha} \\
    \ket{\alpha,-\alpha} &\rightarrow \ket{-\alpha,\alpha} \\
    \ket{-\alpha,\alpha} &\rightarrow \ket{\alpha,-\alpha}. 
\end{align*}

Another potential solution to build a logical Toffoli circuit compatible with a 2D nearest neighbours architecture without using SWAP gates is to use gate teleportation techniques of ~\cite{Gottesman1999}. The implementation of a logical Toffoli gate can be done using transversal logical Clifford operations, and an additional ancillary system of three logical qubits prepared in a special state called the ``Toffoli magic state''. The bottleneck of this approach is the preparation of this magic state with arbitrarily high fidelity. By exploiting the specific structure of noise in  cat qubits, it might be possible to construct a fault-tolerant non-Clifford gate with a similar overhead to the  circuits presented in this paper, while being compatible with a 2D local architecture. \jg{This alternative solution has been thoroughly exposed and quantified in a recent work that was made available} \mm{after the preparation of this manuscript \cite{Chamberland2020_AWS}}.

\section{\label{sec:conclusion}Conclusion}

The realization of high-fidelity quantum operations for large scale quantum processors will most likely involve quantum error correction to achieve fault-tolerance. While this fault-tolerance usually comes at the expense of an important resource overhead, it is pressing to find shortcuts that reduce the overhead requirements to what is achievable in near-term experiments. In this paper, by performing numerical simulations with a realistic circuit-level error model, we showed that the repetition cat qubit  is a serious candidate towards achieving a universal gate set with very low error rates while maintaining the physical cost at a minimum. Similarly to the Bacon-Shor codes, the repetition cat qubit scheme does not possess an accuracy threshold, yet it exhibits a pseudo-threshold in an experimentally realistic regime. In this regime, assuming a physical error probability of $1\%$, a quantum memory with a logical error probability of $10^{-10}$ can be realized using 70 physical cat qubits (twice as much including ancilla cat qubits) and an average number of $\bar n = 15$ photons per mode. The logical gates that are transversal on the repetition code, such as the preparation of the $\ket\pm_L $ states, the measurement of the $X_L$ operator or the CNOT gate, have a similar performance and overhead. The gate that completes the universal set of logical gates is the Toffoli gate, and cannot be implemented transversally. We proposed two different circuits that realize this gate fault-tolerantly. Assuming a physical error probability of $1\%$ per physical Toffoli gate, the first circuit achieves a logical Toffoli error probability of $10^{-10}$ using $9 \times 60 = 540$ cat qubits per logical qubit (1080 including ancillae) and one level of code concatenation. The second circuit is fault-tolerant even without concatenation, by using a tailored error detection protocol preventing both the propagation of errors due to the non-transversality and the accumulation of non-propagating errors. This protocol, involving a more complex circuit, requires the preparation of special ancillary states, but results in an overall reduction of the overhead and achieves a logical error rate of  $10^{-10}$ with only $90$ cat qubits per logical qubit (180 including ancillae). A major obstacle to overcome in this approach, is the requirement of all-to-all couplings  between the cat qubits to implement the logical Toffoli gate. Although it is possible to simplify the connectivity graph by adding SWAP gates, which can themselves be implemented in a bias-preserving manner as required by the construction, the conception of an optimal 2D architecture for a large scale quantum computer based on repetition cat qubits is still under investigation.

\begin{figure*}[t!]
\includegraphics[width=\textwidth]{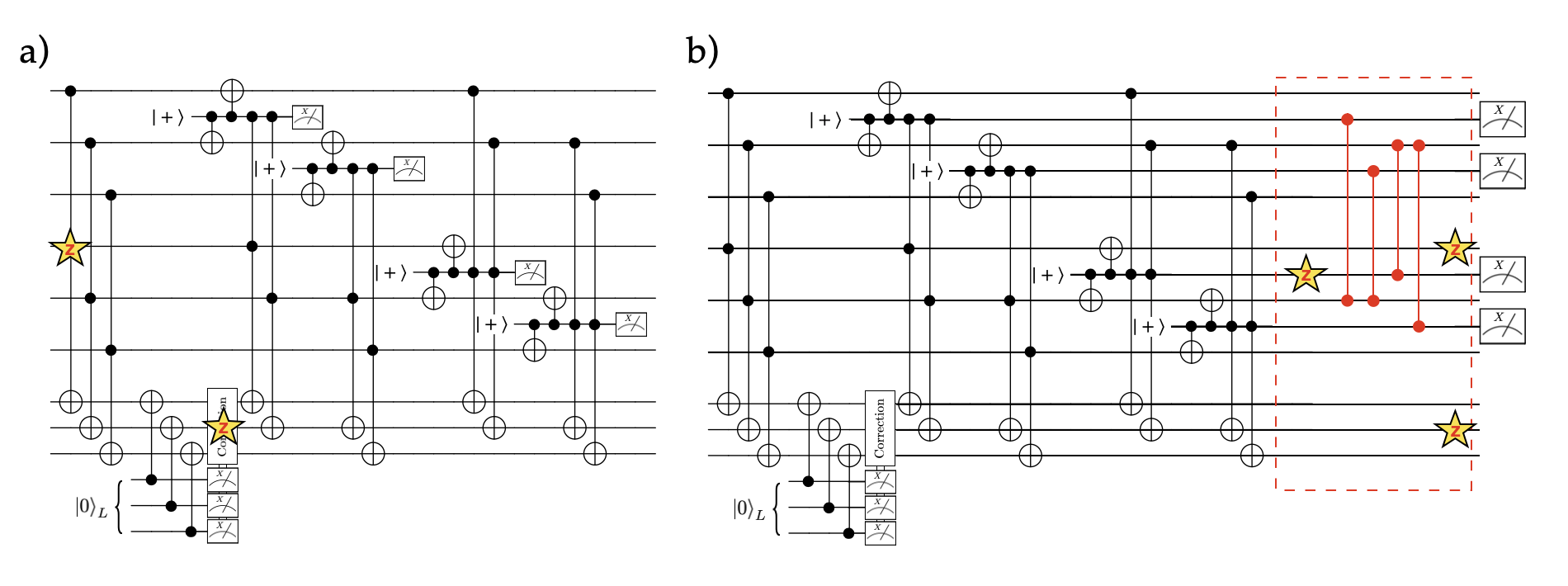}
\caption{Noisy piece of a Toffoli circuit (a) where the errors are drawn randomly from the errors models of Table \ref{tab:error_models}. Here, the first physical Toffoli produces a $Z$ error on the second qubit of the second control block and the Steane error correction produced a $Z$ error on the second qubit of the target block, depicted by the stars. This noisy circuit is equivalent to the one depicted in (b), where the Toffoli circuit is now perfect (error less) and is followed by a circuit containing the noise (in the red dashed box), but composed exclusively of Clifford gates.
\label{fig:noisy_toffoli}}
\end{figure*}

\begin{acknowledgments}
We thank Alain Sarlette and Christophe Vuillot for  fruitful discussions and for their comments on this manuscript, and Donatas Simeliunas for helping with the numerical simulations.
This work has been supported by Region Ile-de-France in the framework of DIM SIRTEQ.  
\end{acknowledgments}

\appendix
\section{\label{appendixA}Efficient simulation of Toffoli circuits}

The simulation of non-Clifford circuits is classically hard. In our case, however, there are a few specific features that enable us to perform the Monte Carlo simulations of all the non-Clifford circuits presented in this work in a classically efficient manner. The first important thing to note is that while the circuits contain non-Clifford Toffoli gates, the propagation of errors in the circuits can never produce non-Clifford errors, as would be the case in all generality. To see this, recall from the error models of Table \ref{tab:error_models} that all the 'bare' errors produced by any gates in the Toffoli circuits are of the following form: a Pauli $Z$ error on any data qubit of any logical block, a Pauli correlated $Z_1Z_2$ error on any two pair of data qubits of the two control blocks, a controlled-phase gate $\text{CZ}_{12}$ between any two data qubit of the control blocks, or a correlated controlled-phase gate and Z erorr $\text{CZ}_{12}Z_3$ on two data qubit of the control block and a data qubit of the target block.
Now, note that none of these errors can become non-Clifford through the Toffoli circuits: the Pauli $Z$ errors of the control blocks commute with the Toffoli gates, while a $Z$ error on the target block evolves through a Toffoli gate as a $Z$ error on the target block together with a controlled-phase error between the two controls:
$$
\text{CCX}_{1,2,3} \times Z_3 = \text{CZ}_{1,2} Z_3 \times \text{CCX}_{1,2,3}.
$$
Thus, the only Clifford error that can ever appear anywhere in the circuits is a controlled-phase between any pair of two qubits of the controls blocks, either produced by a Toffoli gate error or by propagation of a $Z$ error on the target block through another Toffoli gate. Once these errors appear, however, they can never propagate further to non-Clifford errors as a controlled-phase gate on the control qubits of a Toffoli gate commutes with the Toffoli gate:
$$
\text{CCX}_{1,2,3} \times \text{CZ}_{1,2} = \text{CZ}_{1,2} \times \text{CCX}_{1,2,3}.
$$
Thus, several specific ingredients ensure that all the errors we deal with are at most Clifford: the fact that the gates error models are biased, such that we deal only with phase-flip types of errors, the fact that the gates are all bias-preserving, thus preserving the phase-flip nature of errors, and finally the fact that we only use Toffoli gate as non-Clifford resource where the target qubit of any physical Toffoli gate always belong to the same codeblock. Indeed, if we were to use two Toffoli gate in a circuit with the target qubits belonging to two different logical blocks, then a $CZ$ error produced by the first Toffoli gate could evolve to a non-Clifford $CCZ$ error by propagation through the second Toffoli:
$$
\text{CCX}_{1,2,3} \times \text{CZ}_{1',3} = \text{CCZ}_{1,1',2} \times \text{CZ}_{1',3} \times \text{CCX}_{1,2,3}.
$$
With these facts in mind, we now detail how the logical error probability of a noisy Toffoli circuit can be simulated efficiently. The first step is to roll a dice to determine the locations and nature of the errors, according to the errors models described above. Then, the errors are propagated through the gates circuit up until a point where they meet a measurement. At this point, the circuit has been decomposed in two different circuits: the first one is perfect, and contains the non-Clifford Toffoli gates, and it is followed by a second one that consists of the errors only, and contains exclusively Clifford operations. We depict in Fig.~\ref{fig:noisy_toffoli} one example of this circuit decomposition, for the first piece of the circuit of Fig.~\ref{fig:FT_ToffoliCircuit}. The left hand-side of the circuit, which is perfect, is non-Clifford but its effect on the value of the stabilizers is trivial. Actually, since the operators that we measure are ``compatible'' with the Toffoli pieces of the circuit (see Section~\ref{subsec:FT Toffoli}), in the absence of errors, the results of the measurement of these operators is $+1$ with unit probability. The only thing that needs to be simulated numerically to get the correct probability distribution for the measurement outcomes is the effect of the error circuit (red box of Fig.\ref{fig:noisy_toffoli} (b)) on the measurement results and the remaining errors after the measurements have been executed. However, since this error circuit is Clifford, it can be efficiently simulated using the CHP algorithm~\cite{Aaronson2004}. Note that the errors on the target block are always simple Pauli $Z$ errors, even after propagation through any gate of the circuit. Thus, the logical error rate of the target block can actually be simulated separately from the rest, using a simple array of 0, 1 integers.

\bibliography{bibliography}

\end{document}